\documentclass[aps,prl,superscriptaddress,reprint,longbibliography]{revtex4-2}
\usepackage{amsmath}        
\usepackage{amsfonts}
\usepackage{graphicx}
\usepackage{hyperref}
\usepackage{xcolor,soul}
\begin{abstract}
\end{abstract}
                  
\begin{document}
\title{A superconductor free of quasiparticles for seconds}

\author{E. T. Mannila}
\email[]{elsa.mannila@aalto.fi}
\altaffiliation{Present address: VTT Technical Research Centre of Finland Ltd, QTF Centre of Excellence, P.O. Box 1000, FI-02044 VTT, Finland}
\affiliation{QTF Centre of Excellence, Department of Applied Physics, Aalto University, FI-00076 Aalto, Finland}

\author{P. Samuelsson}
\affiliation{Physics Department and NanoLund, Lund University, Box 118, 22100 Lund, Sweden}

\author{S. Simbierowicz}
\altaffiliation{Present address: Bluefors Oy, Arinatie 10, 00370 Helsinki, Finland}
\affiliation{VTT Technical Research Centre of Finland Ltd, QTF Centre of Excellence, P.O. Box 1000, FI-02044 VTT, Finland}

\author{J. T. Peltonen}
\affiliation{QTF Centre of Excellence, Department of Applied Physics, Aalto University, FI-00076 Aalto, Finland}

\author{V. Vesterinen}
\affiliation{VTT Technical Research Centre of Finland Ltd, QTF Centre of Excellence, P.O. Box 1000, FI-02044 VTT, Finland}

\author{L. Gr{\"o}nberg}
\affiliation{VTT Technical Research Centre of Finland Ltd, QTF Centre of Excellence, P.O. Box 1000, FI-02044 VTT, Finland}

\author{J. Hassel}
\altaffiliation{Present address: IQM, Espoo, Finland}
\affiliation{VTT Technical Research Centre of Finland Ltd, QTF Centre of Excellence, P.O. Box 1000, FI-02044 VTT, Finland}

\author{V. F. Maisi}
\affiliation{Physics Department and NanoLund, Lund University, Box 118, 22100 Lund, Sweden}

\author{J. P. Pekola}
\affiliation{QTF Centre of Excellence, Department of Applied Physics, Aalto University, FI-00076 Aalto, Finland}

\date{\today}
\maketitle

\textbf{
Superconducting devices, based on the Cooper pairing of electrons, play an important role in existing and emergent technologies, ranging from radiation detectors \cite{day2003broadband, echternach2018single} to quantum computers \cite{kjaergaard2020superconducting}. Their performance is limited by spurious quasiparticle excitations formed from broken Cooper pairs \cite{aumentado2004nonequilibrium, shaw2008kinetics, devisser2011number,martinis2009energy, catelani2011quasiparticle, pop2014coherent, wang2014measurement, patel2017phononmediated, gustavsson2016suppressing}. Efforts to achieve ultra-low quasiparticle densities have reached time-averaged numbers of excitations on the order of one in state-of-the-art devices \cite{ferguson2008quasiparticle, higginbotham2015parity, vool2014nonpoissonian, gustavsson2016suppressing, echternach2018single}. However, the dynamics of the quasiparticle population as well as the time scales for adding and removing individual excitations remain largely unexplored. Here, we experimentally demonstrate a superconductor completely free of quasiparticles for periods lasting up to seconds. We monitor the quasiparticle number on a mesoscopic superconductor in real time by measuring the charge tunneling to a normal metal contact. Quiet, excitation-free periods are interrupted by random-in-time Cooper pair breaking events, followed by a burst of charge tunneling within a millisecond. Our results demonstrate the possibility of operating devices without quasiparticles with potentially improved performance. In addition, our experiment probes the origins of nonequilibrium quasiparticles in our device; the decay of the Cooper pair breaking rate over several weeks following the initial cooldown rules out processes arising from cosmic or long-lived radioactive sources \cite{vepsalainen2020impact, cardani2020reducing, wilen2020correlated,martinis2020saving}.
}

The Bardeen, Cooper and Schrieffer theory of superconductivity predicts that the number of quasiparticle excitations should be exponentially small at temperatures low compared to the superconducting gap $\Delta$ (divided by Boltzmann's constant $k_\text{B}$). Nevertheless, it is experimentally well established that a residual population persists down to the lowest temperatures \cite{aumentado2004nonequilibrium, shaw2008kinetics, devisser2011number}, attributed to Cooper pair breaking in the superconductor due to non-thermal processes. 
Residual quasiparticles are detrimental to the performance of many superconducting devices, 
ranging from setting the fundamental limit for the sensitivity of kinetic inductance detectors \cite{day2003broadband,devisser2011number} 
to arguably limiting the coherence times of superconducting qubits to the millisecond range \cite{martinis2009energy, catelani2011quasiparticle, vepsalainen2020impact}. 
Correspondingly, their origins and dynamics as well as mitigation schemes have been studied extensively 
 \cite{ferguson2008quasiparticle, pop2014coherent, wang2014measurement, patel2017phononmediated, gustavsson2016suppressing}. 

In several state-of-the art devices, the observed ultra-low quasiparticle densities correspond to only a few excitations within the entire device \cite{ferguson2008quasiparticle, vool2014nonpoissonian, higginbotham2015parity, gustavsson2016suppressing, echternach2018single}.
In this few-particle regime, the number of quasiparticles fluctuates strongly in time \cite{wilson2001timeresolved, devisser2011number, vool2014nonpoissonian, gustavsson2016suppressing, echternach2018single}, potentially leaving the device free of them for certain periods. Identifying such periods would open up possibilities for operation of superconducting devices in the absence of quasiparticles. However, measuring their absolute number in real time is challenging, mainly because breaking of Cooper pairs does not change the total charge. In fact, there are only a few time-resolved experimental investigations of the dynamics in this few-particle regime \cite{ferguson2008quasiparticle, shaw2008kinetics, vool2014nonpoissonian, lambert2014experimental, echternach2018single, serniak2019direct}, and none where quasiparticle-free periods have been identified.

\begin{figure*}[htb]
\includegraphics[width=\textwidth]{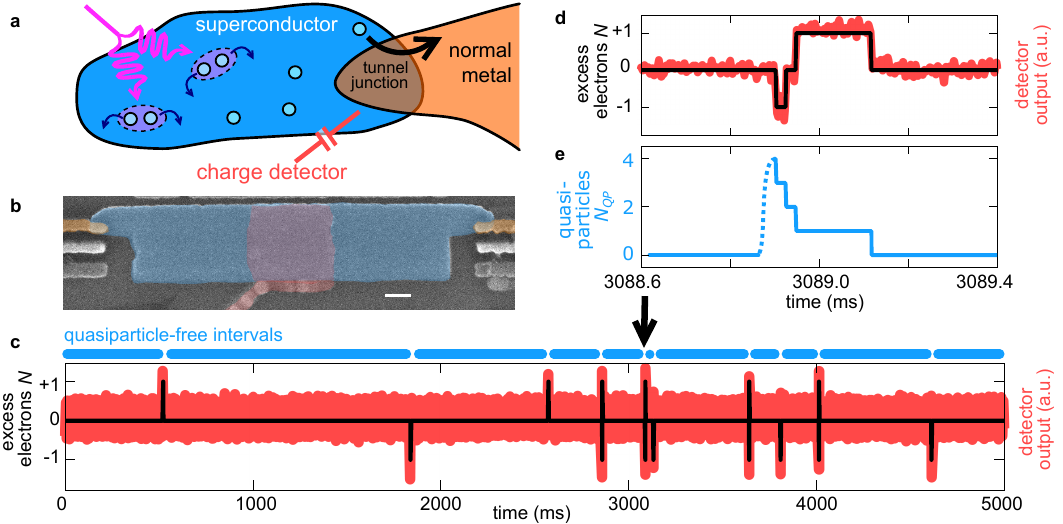}
\caption{\label{fig:schematic}
{\bf Real-time monitoring of the number of quasiparticles on a superconducting island via charge tunneling.}
a, Sketch of experiment. Pair-breaking radiation (pink arrow) is absorbed in a micron-scale superconducting aluminum island (blue), breaking one or more Cooper pairs and producing quasiparticle excitations (light blue circles). The quasiparticles tunnel out to the normal metallic copper leads (orange) via a tunnel junction as either electrons or holes. The tunneling events are measured in real time with a capacitively coupled charge detector (see Methods). b, False-color scanning electron micrograph of the device, with colors as in a. As no voltage bias is applied across the device, the two tunnel junctions act as a single junction. Scale bar, 200 nm. c, Time trace of charge detector output (red line) and the corresponding charge state (black line), $N=0,\pm 1$ excess electrons, of the superconducting island, measured at a refrigerator base temperature of $20$ mK. On top, quasiparticle-free periods are denoted by thick blue lines. d, Zoom in of time trace in c at location indicated by arrow, showing a burst of four single-electron tunneling events ($N$ changing by $\pm 1$). e, Number of quasiparticles $N_{\text{QP}}$ on the island as a function of time, inferred from d. The initial Cooper pair breaking event, occuring on average within 30 $\mu$s before the first tunneling event, is not directly observed (here indicated by a dashed line).}
\end{figure*}

Here, we show experimentally a mesoscopic superconducting island that is free of quasiparticles for time periods reaching seconds, five orders of magnitude longer than typical operation times of superconducting qubits \cite{kjaergaard2020superconducting}. The instantaneous number of quasiparticles is reconstructed from real-time measurements of single-electron tunneling from the island to a normal metal contact, as illustrated in Fig. \ref{fig:schematic}.  We observe quiet periods, with no tunneling events and no quasiparticles on the island, for 99.97\% of the time, interrupted by Cooper pair breaking events, followed by bursts of tunneling within a millisecond. The pair breaking events occur randomly in time, with an average rate 1.5~Hz and a probability 
$\propto \exp(-0.9 N_\text{pair})$ to break $N_\text{pair}$ Cooper pairs.
That is, in 40\% of the events, more than one Cooper pair is broken.

Due to the real time probing of the single-electron tunneling, three orders of magnitude faster than the Cooper pair breaking, we are able not only to measure the decay dynamics of individual quasiparticles but also to indentify the pair breaking events as well the number of quasiparticles created per event. This goes well beyond previous experiments on time-resolved quasiparticle tunneling in superconducting structures \cite{shaw2008kinetics, lambert2014experimental, vanwoerkom2015one, serniak2019direct, echternach2018single, hays2018direct} 
or experiments with intentional injection of a large number of quasiparticles \cite{wang2014measurement, patel2017phononmediated}. The very good agreement between the measured dynamics and a rate equation model for the quasiparticle number allows for an unambiguous identification of the excitation-free periods. Our findings create opportunities for operating superconducting devices while simultaneously monitoring the instantaneous quasiparticle number, allowing for identifying periods for quasiparticle-free operation. This would enable testing predictions of improved device performance in the absence of quasiparticles \cite{martinis2009energy, catelani2011quasiparticle, karzig2020quasiparticle}. In addition, the statistics of number of quasiparticles created might provide additional information on the origin of the pair-breaking processes. Here, we observe that the rate of Cooper pair breaking events decreased by a factor of 4 over a period of weeks. This rules out commonly suggested sources of nonequilibrium quasiparticles, such as insufficient shielding against stray light \cite{serniak2019direct, barends2011minimizing}
or ionizing radiation \cite{vepsalainen2020impact, cardani2020reducing, wilen2020correlated, martinis2020saving}.   

We monitor the charge on an aluminum superconducting island, sketched in Fig. \ref{fig:schematic}a and shown in Fig. \ref{fig:schematic}b, with volume $V=2$ $\mu$m~$\times~0.5~\mu$m~$\times~35~$nm, charging energy $E_C$ $\approx 90~\mu$eV and superconducting gap $\Delta$ $\approx 220~\mu$eV. The island is coupled to normal metal copper leads via an insulating aluminum oxide tunneling barrier. Measurements were done in a dilution refrigerator with a base temperature of 20 mK at a normal metal electron temperature $T_N \approx$ 100 mK, obtained by fitting the temperature dependence of two-electron Andreev tunneling rates (Supplementary Note 6). Coulomb blockade limits the observed charge states to $N=0, \pm 1$ excess electrons on the island. The charge detector is a capacitively coupled radio-frequency single-electron transistor operating at 580 MHz whose output is amplified with a Josephson parametric amplifier with an ultralow added noise of about 100 mK, enabling single-shot charge readout with a high signal-to-noise ratio in a few $\mu$s. Details on the sample fabrication, measurements and device characterization are given in Methods and Supplementary Notes 1, 2, 3 and 6. 

\begin{figure}[tb]
\includegraphics{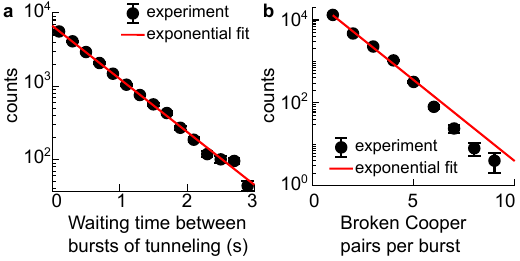}
\caption{\label{fig:events}
{\bf Statistics of Cooper pair breaking events.} 
a, Lengths of the quiet periods between bursts of tunneling. The experimental data (circles), binned into intervals of equal length 0.2~s, is fitted with an exponential curve (solid line). The corresponding Poissonian waiting time distribution describes random-in-time bursts with an average burst rate $\Gamma_{\text{burst}} \approx 1.7$~Hz and average length of quiet period $\approx 0.4$~s. Note that $\Gamma_{\text{burst}}$ overestimates the true burst rate, $1.5$~Hz, due to finite length of time traces (Supplementary Note 4). b, Counts of number of quasiparticle pairs $N_{\text{pair}}$ ($N_{\text{QP}}=2N_{\text{pair}}$) generated per event. Experimental data (circles) is obtained by counting the number of single electron tunneling events within each burst, see Fig. \ref{fig:schematic}c,d. The solid line is an exponential fit $\sim \exp (-\lambda N_{\text{pair}} )$, weighted by the counts, with $\lambda=0.9 \pm 0.1$. Error bars in both panels represent statistical uncertainty due to the finite number of events per bin. 
}
\end{figure}

We measure 5 s time traces of the detector output. A representative trace is shown in Fig. \ref{fig:schematic}c. Quiet periods up to seconds long, with no tunneling events and zero excess electrons, are interrupted by bursts of two or more one- or two-electron tunneling events between the charge states $N=0,\pm 1$ within a millisecond, see Fig. \ref{fig:schematic}d. We collect statistics from 4.6~h of time traces, containing over $2 \times 10^4$ bursts of tunneling. In Fig. \ref{fig:events}a we show that the waiting times between the bursts of tunneling follow an exponential distribution, characterized by a rate $\Gamma_{\text{burst}} \approx 1.5$~Hz. This implies that the bursts occur independently, in a Poissonian process. Within the bursts of tunneling, the average rates of single-electron tunneling are $\Gamma_{0 \rightarrow \pm 1} \approx$ 20~kHz and $\Gamma_{\pm 1 \rightarrow 0} \approx$ 11~kHz. Here rate subscripts denote the initial and final charge states. The tunneling rates are thus several orders of magnitude higher than the rate at which the bursts occur. 

To infer the instantaneous number of quasiparticles $N_{\text{QP}}$ from the time traces we make a number of observations (see supplement for details): (i) The single-electron tunneling rates $\Gamma_{0 \rightarrow \pm 1}$ and $\Gamma_{\pm 1 \rightarrow 0}$ are constant over a range of the gate offset $n_g$, as expected for quasiparticle excitations \cite{saira2012vanishing}.  (ii) The number of single-electron tunneling events within a burst is even. (iii) The two-electron events are due to Andreev tunneling which does not change $N_{\text{QP}}$. (iv) Based on the measured values of $T_N$, $\Delta$ and $E_C$ and further analysis we conclude that the rate for quasiparticles tunneling back into the superconductor from the normal metal leads 
is much smaller than both single- and two-electron tunneling rates. (v) Moreover, based on earlier experiments \cite{maisi2013excitation, wang2014measurement}, the recombination of two quasiparticles into a Cooper pair is more than an order of magnitude slower than single-electron tunneling. Taken together, these observations imply the following compelling, simple physical picture of the quasiparticle dynamics, illustrated in Fig. \ref{fig:schematic}e: A burst of tunneling is initiated by the breaking of $N_{\text{pair}}$ Cooper pairs, creating $N_{\text{QP}}=2N_{\text{pair}}$ quasiparticles on the island. They tunnel out to the normal leads as electrons or holes via single-electron tunneling events, which decrease the number of quasiparticles one by one, $N_{\text{QP}}\rightarrow N_{\text{QP}}-1$ as the charge state $N$ changes by $\pm1$. After the last quasiparticle has tunneled out, within a millisecond from the start of the burst, the superconducting island is \textit{completely free of quasiparticles} until the next burst occurs, on average for 0.4 seconds. Therefore counting the number of events changing $N$ yields directly $N_\text{QP}$, which also allows us to infer  $N_\text{pair}$ for each burst. 

\begin{figure*}[t]
\includegraphics{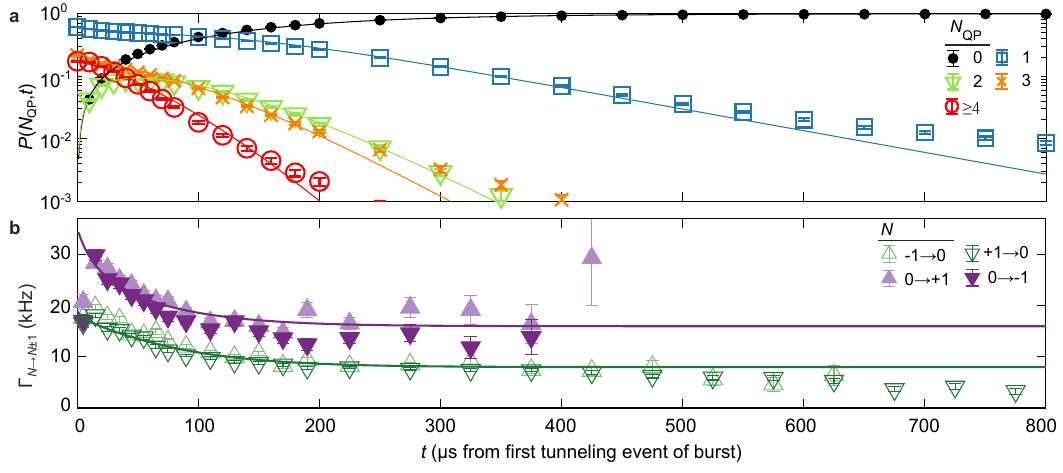}
\caption{\label{fig:within}
{\bf Dynamics of quasiparticle relaxation within a burst.}
a, The probability $P(N_{\text{QP}},t)$ to have $N_{\text{QP}}$ quasiparticles present on the island at time $t$, for $N_{\text{QP}}=0,1,2,3$ and $N_{\text{QP}}\geq 4$. Already at $t=150$~$\mu$s the probability that the burst has ended, $P(N_{\text{QP}}=0)$, is above one half. The theoretical model (solid lines), see text and Methods, assumes quasiparticle relaxation by tunneling only, and uses the measured distribution of broken Cooper pairs, Fig. \ref{fig:events}b, as an initial condition. 
b, Effective single-electron tunneling rates $\Gamma_{N \rightarrow N \pm 1}(t)$ as a function of time $t$ from the first tunneling event of a burst. After an initial decay, the rates saturate to minimum values corresponding to tunneling rates for one (two) quasiparticles on the island for $N$ odd (even). The dip in the rates at short times ($t<20$ $\mu$s) is due to finite bandwidth of the experiment, as discussed in Methods. Data is shown only if there are at least 10 bursts with events in the corresponding bin.
In both panels a and b, experimental data (symbols) are obtained from averaging over $2 \times 10^4$ bursts, and error bars reflect statistical uncertainty due to the finite number of events per time bin. For the theoretical model (solid lines) $\Gamma_{\text{QP}}=8.0$~kHz is used.}
\end{figure*}

The statistical distribution of $N_\text{pair}$, shown in Fig. \ref{fig:events}b, is well described by an exponential $\propto \exp(-\lambda N_\text{pair})$ with $\lambda \approx 0.9 \pm 0.1$. In 40\% of the bursts, more than one Cooper pair was thus initially broken. Further insight into the dynamics of quasiparticle number relaxation is obtained from the probability $P(N_{\text{QP}},t)$ of having $N_\text{QP}$ quasiparticles at a time $t$ after the first tunneling event in the burst. The probabilities for $N_{\text{QP}}=0$ to $3$ as well as $N_{\text{QP}} \geq 4$ are shown in Fig. \ref{fig:within}a. Already at $t=0.5$ ms, the probability for having any quasiparticles remaining is below 10$\%$. At short times $P(N_{\text{QP}},t)$ increases (decreases) for $N_{\text{QP}}$ even (odd) as a function of $t$, a consequence of having an odd number of quasiparticles on the island directly after the first tunneling event at $t=0$. 

Figure \ref{fig:within}a also shows that $P(N_{\text{QP}},t)$ decays faster for increasing $N_{\text{QP}}$. This is consistent with the prediction that the single electron tunneling rate for $N_{\text{QP}}$ quasiparticles on the island is proportional to $N_{\text{QP}}$  \cite{saira2012vanishing}. It is further reflected in the time-dependent rates $\Gamma_{0 \rightarrow \pm 1}(t)$ and $\Gamma_{\pm 1 \rightarrow 0}(t)$ for single electron tunneling shown in Fig. \ref{fig:within}b, evaluated in a short interval around time $t$ within the bursts. In particular, we see that the long time ($t > 200\ \mu$s) rate from the state $N=0$, $\Gamma_{0 \rightarrow \pm 1}(t) \approx 15$~kHz, is twice the rate from $N=\pm1$, $\Gamma_{\pm 1 \rightarrow 0}(t) \approx 8$~kHz, since the even state has two quasiparticles before the end of the burst compared to one quasiparticle for odd states.  
Furthermore, the initial decay is consistent with the picture that initially more than two quasiparticles may be created. The timescale for the decay arises from the quasiparticle tunneling rate and is thus of the order of 100~$\mu$s.
The measured single-quasiparticle rate $\Gamma_{\text{QP}} = \Gamma_{\pm 1 \rightarrow 0}(t) = 8~$kHz is close to the estimate 17 kHz obtained based on other experiments (Supplementary Note 6). % Exact values from weighted averages are 15.3 kHz and 7.6 kHz.

The dynamics of $N_\text{QP}$ within a burst, based on the simple physical picture, are modelled by a rate equation for $P(N_{\text{QP}},t)$, see Methods. As the initial condition, we have the experimentally determined distribution $P(N_{\text{QP}}) \propto \exp(-0.9 N_{\text{pair}}) = \exp(-0.45 N_{\text{QP}})$ at the start of the burst, see Fig. \ref{fig:events}. Only tunneling out of quasiparticles is accounted for, making $\Gamma_{\text{QP}}$ the only free parameter of the model. From Fig. \ref{fig:within}a we see that the model reproduces the experimental results well for all times within the burst and over three orders of magnitude of $P(N_{\text{QP}},t)$, for all $N_{\text{QP}}$ shown. The best fit is obtained for $\Gamma_{\text{QP}}=8.0$~kHz, in agreement with the measurent in Fig. \ref{fig:within}b. Moreover, in Fig. \ref{fig:within}b the results from the model for the time-dependent rates $\Gamma_{0 \rightarrow \pm 1}(t)$ and $\Gamma_{\pm 1 \rightarrow 0}(t)$ are plotted together with the experimental data. Both the observed initial decay and the long time saturation are well captured by the model. 
The model also allows us (see Methods) to obtain an accurate estimate of the time-averaged number of quasiparticles on the island $\langle N_{\text{QP}} \rangle=2.5\, \Gamma_{\text{burst}}/\Gamma_{\text{QP}} =4.7 \times 10^{-4}$, measured 130 days after the start of the cooldown. The corresponding quasiparticle density is $n_{\text{QP}}$ = $\langle N_{\text{QP}} \rangle / V = $ 0.013 quasiparticles/$\mu$m$^{3}$ or $x_{\text{QP}}=\langle N_{\text{QP}} \rangle/(D(E_\text F)\Delta V) \approx 2\times 10^{-9}$, if normalized by the Cooper pair density $\Delta D(E_\text F)$, with $D(E_\text F)=2.15 \times 10^{47}$~J$^{-1}$m$^{-3}$ the density of states. This is comparable to the lowest measured values $x_\text{QP} \sim 10^{-9}$ reported in the literature \cite{serniak2019direct} and below the bound $x_\text{QP} \approx 7 \times 10^{-9}$ estimated due to the ionizing radiation background in Ref. \cite{vepsalainen2020impact}. 

\begin{figure}
\includegraphics{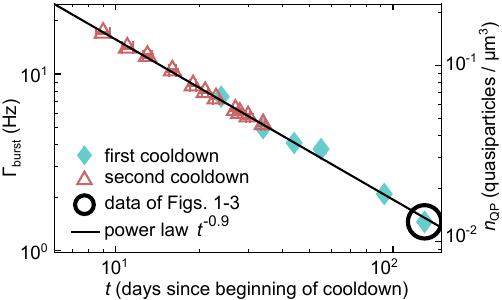}
\caption{\label{fig:cooldown}
\textbf{Decay of burst rate over time.}
The observed rate of Cooper pair breaking events leading to bursts of tunneling $\Gamma_{\text{burst}}$ is shown as a function of time $t$ after the start of the first (filled diamonds) and second (open triangles) cooldown. A power law fit $\Gamma_{\text{burst}} \propto t^{-0.9}$ is shown by a solid line. The right y-axis shows the corresponding time-averaged quasiparticle density $n_{\text{QP}}$, proportional to $\Gamma_{\text{burst}}$ (see text) since the distribution of quasiparticles generated per burst as well as the single electron tunneling rate were found to be constant over the entire cooldown period (Supplementary Note 7). A circle indicates the data used in Figs. 1-3, and error bars indicate statistical uncertainty as well uncertainty in the timing from the start of the cooldown. 
}
\end{figure}

In our experiment, the rate of the bursts $\Gamma_{\text{burst}}$ decreased over time from the start of a cooldown on timescales of weeks following a power law $\Gamma_{\text{burst}} \propto t^{-0.9}$, where $t$ is the time after the sample reached temperatures below 77~K. This was observed reproducibly in two subsequent cooldowns of the same sample, as shown in Fig. \ref{fig:cooldown}. 
The observed long time decay of $\Gamma_{\text{burst}}$ rules out a number of observed or suggested sources of quasiparticles being dominant in our system. In particular, external infrared or microwave photons \cite{barends2011minimizing, serniak2019direct}  and environmental radioactivity and cosmic rays \cite{cardani2020reducing, vepsalainen2020impact, wilen2020correlated, martinis2020saving} 
are all expected to be stationary sources of quasiparticles. Moreover, the distribution of Cooper pairs broken per burst did not change during the cooldown (Supplementary Figures S10-S12), suggesting that the same source of quasiparticles was dominant throughout the experiment. Although heat leaks decreasing over timescales of weeks are well known in experiments at low temperatures \cite{pobell}, we cannot certify what in our experiment is the mechanism of Cooper pair breaking. 

The real-time monitoring of the single-electron tunneling (see Fig. \ref{fig:schematic}) reveals the dynamics of the charge state of the island charge $N$, and hence the parity, during a burst. On a typical time scale $1/\Gamma_{\text{QP}}$, the charge changes from $N=0$ to $N=\pm 1$ and back, a number of times. However, for less than 1\% of the bursts, a distinctly different dynamics is observed; after one or several tunnel events the charge state appears to get stuck at $N=\pm 1$ for a period of time much longer than $1/\Gamma_{\text{QP}}$, before changing back to $N=0$ via a number of additional tunnel events (see Supplementary Note 8 and Supplementary Figure S13). We attribute the switching between different type of charge dynamics, not described by our simple model, to trapping and subsequent escape of a quasiparticle in a localized state. Within this interpretation the quasiparticle trapping and escape shifts the offset charge $n_g$ by one electron, thereby altering the tunnel dynamics but not the number of charges $N$ of the island. During the first cooldown, the trapping events occur on a timescale of hours, while in the second cooldown, they have a rate of the order of 1 Hz. The rates $\Gamma_\text{burst}$ and $\Gamma_{\text{QP}}$, however, do not change between cooldowns. Taken together, this illustrates that the dynamics of the island charge state, or parity, in our experiment is governed by a number of different physical processes and cannot be captured by a single time scale.   

In conclusion, we have investigated the dynamics of the number of quasiparticles on a superconducting island via real time monitoring of single electron tunneling. We demonstrate that a mesoscopic superconducting island can remain completely free from quasiparticles for time periods up to seconds. These findings open a new route to identifying periods of quasiparticle free operation of superconducting devices, of key interest for e.g. superconducting quantum computation. Our approach, giving access to the number of Cooper pairs broken in a given event, also provides novel information on the properties of the source of quasiparticles, which is presently an actively investigated and debated topic \cite{cardani2020reducing, vepsalainen2020impact, wilen2020correlated, martinis2020saving}. 
Furthermore, our device could potentially be adapted to operate as an energy-resolving single-photon detector in the terahertz range similar to the quantum capacitance detector \cite{echternach2018single}.

\section{Acknowledgements}

We acknowledge useful discussions with O. Maillet and support in shielding solutions from J. Ala-Heikkilä. 
We thank Janne Lehtinen and Mika Prunnila from VTT Technical Research Center of Finland Ltd also involved in the JPA development.
This work was performed as  part of the Academy of Finland Centre of Excellence program, projects 312057 (E.T.M., J.T.P and J.P.P.), 312059 (S.S., V.V., and J.H.) and 312294 (L.G.). 
We acknowledge the provision of facilities and technical support by Aalto University at OtaNano - Micronova Nanofabrication Centre and OtaNano - Low Temperature Laboratory.
E.T.M. and J.P.P. acknowledge financial support from Microsoft. V.V. acknowledges financial support from the Academy of Finland through grant no. 321700. V.F.M. and P.S. were supported by the Swedish Research Council, grant numbers 2019-04111 (V.F.M.) and 2018-03921 (P.S.). V.F.M. acknowledges financial support from the QuantERA project “2D hybrid materials as a platform for topological quantum computing” and NanoLund.

\section{Author contributions}
E.T.M., P.S., V.F.M. and J.P.P. conceived the experiment and model and interpreted the results.
E.T.M. fabricated the sample and performed the experiment with support from J.T.P., and E.T.M. analysed the data. P.S. performed the theoretical modeling. S.S., V.V., L.G. and J.H. provided the Josephson parametric amplifier. E.T.M. and J.T.P. integrated the Josephson parametric amplifier into the setup with assistance from S.S. and V.V. 
The manuscript was written by E.T.M. and P.S. with input from all coauthors.

\section{Competing interests}
The authors declare no competing interests.

%\section{Figures}

\section{Methods}

\subsection{Sample fabrication}
 
The sample is fabricated by standard electron-beam lithography and multiple angle deposition on a silicon substrate. The normal metal leads of the superconducting island are capacitively shunted by a 2 nm / 30 nm / 2 nm thick Ti/Au/Ti ground plane, which helps suppress photon-assisted tunneling events \cite{pekola2010environment}. The charge detector is a normal-metallic single-electron transistor fabricated with the laterally proximitized tunnel junction technique \cite{koski2011laterally}, expected to minimize quasiparticle poisoning due to nonequilibrium phonons \cite{ patel2017phononmediated}.
A 10 $\mu$m long chromium capacitor creates the capacitive coupling between the detector and superconductor island. The ground plane and coupler are insulated from the actual devices by a 40 nm thick aluminum oxide layer grown by atomic layer deposition. Further details on the sample design and fabrication are given in Supplementary Note 1 and Supplementary Figure S1. 

\subsection{Measurements}

The sample is attached to a copper sample holder weighing approximately 0.5 kg with vacuum grease. Aluminum wire bonds connect the sample chip and the resonator chip (see below) to the printed circuit board of the sample stage. The bond wires to one of the bias leads and the gate electrode of the superconducting island had disconnected during the cooldown, but we were able to tune $n_g$ by applying a voltage to the surviving bias lead instead. In the measurements shown in the main text, the gate voltage was held constant at $n_g=0$. The sample holder is closed with a single indium-sealed cap, which is coated by Eccosorb. The sample holder is thermally anchored to the mixing chamber stage of a cryogen-free dilution refrigerator with a base temperature of 20 mK, with the innermost radiation shield attached to the still flange of the refrigerator (temperature between 0.5 K and 1 K).

The low-frequency measurement lines to the sample are filtered with approximately 2 m of Thermocoax \cite{zorin1995thermocoax}. The radio-frequency lines are filtered and attenuated with commercial attenuators and filters at different temperature stages of the refrigerator. We have measured the same sample in two subsequent cooldowns. For the second cooldown, we added home-made Eccosorb filters to the input and output RF lines at the mixing chamber stage, as well as disconnecting a RF gate line not used in this experiment. A detailed wiring schematic is provided in Supplementary Figure S2 and Supplementary Note 2. 

We operate the charge detector as a radio frequency single-electron transistor \cite{schoelkopf1998radiofrequency} at 580 MHz, with details on the detector given in Supplementary Note 3. 
We have verified by changing the probe power that the detector does not cause measurable backaction on the superconducting island in this experiment (Supplementary Figure S3). % Power -50 (first cooldown) or -55 (snd) at ref + 20 from coupler -81 dB attenuator, eccosorb is lossy + 12 dBm probe
The output signal from the detector is amplified by a Josephson parametric amplifier (JPA) with added noise on the order of 100 mK \cite{simbierowicz2018fluxdriven} at the mixing chamber, followed by semiconductor amplifiers at the 2 K stage and at room temperature. The parametric amplifier is similar to the devices presented in Ref. \cite{simbierowicz2018fluxdriven}, but has a higher bandwidth. Operating the JPA in the phase-sensitive mode enables charge sensitivities of $2 \times 10^{-5} e/\sqrt{\text{Hz}}$ and charge detection with a signal-to-noise ratio of 6 in 3~$\mu$s (Supplementary Figure S4). We note that the measured time trace appears noisier in Fig. 1c than in 1d although the same data is shown on both panels, but this is purely a visual effect. The JPA improves the time resolution of the experiment by roughly a factor of 20.

The amplified signal is down-converted and digitized at room temperature with a Aeroflex 3035C vector digitizer. We save both quadratures of the signal in 5 s traces with a sampling rate of 4 MHz, and rotate the signal such that the charge response is along one quadrature only. We then low-pass filter the signal digitally with a cutoff frequency of 150 kHz to increase the signal-to-noise. This post-processing filter sets the bandwidth of our charge detector. We then identify the charge states, bursts of tunneling, and single- and two-electron tunneling events from the filtered time traces as described in Supplementary Note 4 and Supplementary Figures S5 and S6.  

\subsection{Model}

The quasiparticle dynamics is governed by a rate equation for the probability distribution $P(N_{\text{QP}},t)$ of having $N_{\text{QP}}$ quasiparticles on the island at a time $t$ after the first tunneling event in a burst. From the general model, see Supplementary Notes 5 and 6 and Supplementary Figure S7, considering only single-electron tunneling out from the island we have   
\begin{eqnarray}
\frac{dP(N_{\text{QP}},t)}{dt}&=&-\sigma \Gamma_{\text{QP}}N_{\text{QP}}P(N_{\text{QP}},t) \nonumber \\
&+&\bar \sigma \Gamma_{\text{QP}}(N_{\text{QP}}+1)P(N_{\text{QP}}+1,t)
\label{methodrateeq}
\end{eqnarray}
where $\sigma=1, \bar \sigma=2$ ($\sigma=2, \bar \sigma=1$) for $N_{\text{QP}}=1,3,5,..$ ($N_{\text{QP}}=0,2,4,..$). The factors $\sigma$ arise from the fact that starting from the state $N=0$, where $N_{\text{QP}}$ is even, the quasiparticles can tunnel out as both electrons and holes with two possible final states $N=\pm1$. On the other hand, starting from $N=\pm 1$, where $N_{\text{QP}}$ is odd, strong Coulomb interactions restrict the possible final states to $N=0$. With the initial condition given from the observed exponential distribution of number of broken Cooper pairs, see Fig. \ref{fig:events}b, the probability distribution plotted in Fig. \ref{fig:within}a is evaluated numerically. The effective, time dependent single-electron tunneling rates $\Gamma_{\pm 1 \rightarrow 0}(t)$ and $\Gamma_{\pm 1 \rightarrow 0}(t)$, plotted in Fig. \ref{fig:within}b, are obtained from (see Supplementary Note 5 for details)  
\begin{eqnarray}
\Gamma_{N \rightarrow N \pm 1}(t)=\Gamma_{\text{QP}}\frac{\sum_{N_{\text{QP}}} \sigma P(N_{\text{QP}},t) N_{\text{QP}} }{\sum_{N_{\text{QP}}} P(N_{\text{QP}},t)},
\end{eqnarray}
where the sum runs over $N_\text{QP}=2,4,6,...$ for $N=0$ and $N_\text{QP}=1,3,5,...$ for $N=\pm 1$. 

In the text the time-averaged number of quasiparticles $\langle N_{\text{QP}} \rangle$ on the island is discussed. From the probability distribution in Eq. (\ref{methodrateeq})  we can evaluate  $\langle N_{\text{QP}} \rangle$ along the following lines: From the observed distribution of waiting times between Cooper pair breaking events in Fig. \ref{fig:events}a we have that the average burst rate is $\Gamma_{\text{burst}}$. During a burst, the time dependent number of quasiparticles $N_{\text{QP}}(t)$ on the island is given by
\begin{eqnarray}
N_{\text{QP}}(t)=\sum_{N_{\text{QP}}} P(N_{\text{QP}},t) N_{\text{QP}}, ~N_{\text{QP}}=0,1,2,.. 
\end{eqnarray}
The average number $\langle N_{\text{QP}} \rangle$ is then directly given by 
\begin{eqnarray}
\langle N_{\text{QP}} \rangle=\Gamma_{\text{burst}}\int_0^{\infty} N_{\text{QP}}(t) dt=\frac{\Gamma_{\text{burst}}}{\Gamma_{\text{QP}}}c(\lambda)
\end{eqnarray}
where a numerical evaluation gives the coefficient $c(\lambda=0.45)=2.5$, which depends on the initial condition. We note that the $1/\Gamma_{\text{QP}}$ dependence arises since $P(N_{\text{QP}},t)$, from Eq. (\ref{methodrateeq}), depends only on the renormalized time $t\Gamma_{\text{QP}}$. Moreover, the upper limit $\infty$ of the integral can be taken due to the orders of magnitude difference between the burst duration $\sim 1/\Gamma_{\text{QP}}$ and the average time between bursts $1/\Gamma_{\text{burst}}$. We can also estimate the probability of having undetected quasiparticles on the island at any given moment of time within the quiet periods, which is less than $10^{-6}$ based on the probabilities of missing tunneling events due to the finite detector bandwidth and quasiparticles decaying by recombination instead of tunneling (see Supplementary Notes 4 and 6 for details).

\section{Data availability}

Data for figures that support the manuscript are available at https://doi.org/10.5281/zenodo.5574415. All other data that support the findings of this study are available from the corresponding author upon reasonable request.

%\bibliography{phonons}

\begin{thebibliography}{34}%
\makeatletter
\providecommand \@ifxundefined [1]{%
 \@ifx{#1\undefined}
}%
\providecommand \@ifnum [1]{%
 \ifnum #1\expandafter \@firstoftwo
 \else \expandafter \@secondoftwo
 \fi
}%
\providecommand \@ifx [1]{%
 \ifx #1\expandafter \@firstoftwo
 \else \expandafter \@secondoftwo
 \fi
}%
\providecommand \natexlab [1]{#1}%
\providecommand \enquote  [1]{``#1''}%
\providecommand \bibnamefont  [1]{#1}%
\providecommand \bibfnamefont [1]{#1}%
\providecommand \citenamefont [1]{#1}%
\providecommand \href@noop [0]{\@secondoftwo}%
\providecommand \href [0]{\begingroup \@sanitize@url \@href}%
\providecommand \@href[1]{\@@startlink{#1}\@@href}%
\providecommand \@@href[1]{\endgroup#1\@@endlink}%
\providecommand \@sanitize@url [0]{\catcode `\\12\catcode `\$12\catcode
  `\&12\catcode `\#12\catcode `\^12\catcode `\_12\catcode `\%12\relax}%
\providecommand \@@startlink[1]{}%
\providecommand \@@endlink[0]{}%
\providecommand \url  [0]{\begingroup\@sanitize@url \@url }%
\providecommand \@url [1]{\endgroup\@href {#1}{\urlprefix }}%
\providecommand \urlprefix  [0]{URL }%
\providecommand \Eprint [0]{\href }%
\providecommand \doibase [0]{https://doi.org/}%
\providecommand \selectlanguage [0]{\@gobble}%
\providecommand \bibinfo  [0]{\@secondoftwo}%
\providecommand \bibfield  [0]{\@secondoftwo}%
\providecommand \translation [1]{[#1]}%
\providecommand \BibitemOpen [0]{}%
\providecommand \bibitemStop [0]{}%
\providecommand \bibitemNoStop [0]{.\EOS\space}%
\providecommand \EOS [0]{\spacefactor3000\relax}%
\providecommand \BibitemShut  [1]{\csname bibitem#1\endcsname}%
\let\auto@bib@innerbib\@empty
%</preamble>
\bibitem [{\citenamefont {Day}\ \emph {et~al.}(2003)\citenamefont {Day},
  \citenamefont {LeDuc}, \citenamefont {Mazin}, \citenamefont {Vayonakis},\
  and\ \citenamefont {Zmuidzinas}}]{day2003broadband}%
  \BibitemOpen
  \bibfield  {author} {\bibinfo {author} {\bibfnamefont {P.~K.}\ \bibnamefont
  {Day}}, \bibinfo {author} {\bibfnamefont {H.~G.}\ \bibnamefont {LeDuc}},
  \bibinfo {author} {\bibfnamefont {B.~A.}\ \bibnamefont {Mazin}}, \bibinfo
  {author} {\bibfnamefont {A.}~\bibnamefont {Vayonakis}},\ and\ \bibinfo
  {author} {\bibfnamefont {J.}~\bibnamefont {Zmuidzinas}},\ }\bibfield  {title}
  {\bibinfo {title} {A broadband superconducting detector suitable for use in
  large arrays},\ }\href {https://doi.org/10.1038/nature02037} {\bibfield
  {journal} {\bibinfo  {journal} {Nature}\ }\textbf {\bibinfo {volume} {425}},\
  \bibinfo {pages} {817} (\bibinfo {year} {2003})}\BibitemShut {NoStop}%
\bibitem [{\citenamefont {Echternach}\ \emph {et~al.}(2018)\citenamefont
  {Echternach}, \citenamefont {Pepper}, \citenamefont {Reck},\ and\
  \citenamefont {Bradford}}]{echternach2018single}%
  \BibitemOpen
  \bibfield  {author} {\bibinfo {author} {\bibfnamefont {P.~M.}\ \bibnamefont
  {Echternach}}, \bibinfo {author} {\bibfnamefont {B.~J.}\ \bibnamefont
  {Pepper}}, \bibinfo {author} {\bibfnamefont {T.}~\bibnamefont {Reck}},\ and\
  \bibinfo {author} {\bibfnamefont {C.~M.}\ \bibnamefont {Bradford}},\
  }\bibfield  {title} {\bibinfo {title} {Single photon detection of 1.5 {THz}
  radiation with the quantum capacitance detector},\ }\href
  {https://doi.org/10.1038/s41550-017-0294-y} {\bibfield  {journal} {\bibinfo
  {journal} {Nature Astronomy}\ }\textbf {\bibinfo {volume} {2}},\ \bibinfo
  {pages} {90} (\bibinfo {year} {2018})}\BibitemShut {NoStop}%
\bibitem [{\citenamefont {Kjaergaard}\ \emph {et~al.}(2020)\citenamefont
  {Kjaergaard}, \citenamefont {Schwartz}, \citenamefont {Braumüller},
  \citenamefont {Krantz}, \citenamefont {Wang}, \citenamefont {Gustavsson},\
  and\ \citenamefont {Oliver}}]{kjaergaard2020superconducting}%
  \BibitemOpen
  \bibfield  {author} {\bibinfo {author} {\bibfnamefont {M.}~\bibnamefont
  {Kjaergaard}}, \bibinfo {author} {\bibfnamefont {M.~E.}\ \bibnamefont
  {Schwartz}}, \bibinfo {author} {\bibfnamefont {J.}~\bibnamefont
  {Braumüller}}, \bibinfo {author} {\bibfnamefont {P.}~\bibnamefont {Krantz}},
  \bibinfo {author} {\bibfnamefont {J.~I.-J.}\ \bibnamefont {Wang}}, \bibinfo
  {author} {\bibfnamefont {S.}~\bibnamefont {Gustavsson}},\ and\ \bibinfo
  {author} {\bibfnamefont {W.~D.}\ \bibnamefont {Oliver}},\ }\bibfield  {title}
  {\bibinfo {title} {Superconducting qubits: Current state of play},\ }\href
  {https://doi.org/10.1146/annurev-conmatphys-031119-050605} {\bibfield
  {journal} {\bibinfo  {journal} {Annual Review of Condensed Matter Physics}\
  }\textbf {\bibinfo {volume} {11}},\ \bibinfo {pages} {369} (\bibinfo {year}
  {2020})}\BibitemShut {NoStop}%
\bibitem [{\citenamefont {Aumentado}\ \emph {et~al.}(2004)\citenamefont
  {Aumentado}, \citenamefont {Keller}, \citenamefont {Martinis},\ and\
  \citenamefont {Devoret}}]{aumentado2004nonequilibrium}%
  \BibitemOpen
  \bibfield  {author} {\bibinfo {author} {\bibfnamefont {J.}~\bibnamefont
  {Aumentado}}, \bibinfo {author} {\bibfnamefont {M.~W.}\ \bibnamefont
  {Keller}}, \bibinfo {author} {\bibfnamefont {J.~M.}\ \bibnamefont
  {Martinis}},\ and\ \bibinfo {author} {\bibfnamefont {M.~H.}\ \bibnamefont
  {Devoret}},\ }\bibfield  {title} {\bibinfo {title} {Nonequilibrium
  quasiparticles and $2e$ periodicity in single-{Cooper}-pair transistors},\
  }\href {https://doi.org/10.1103/PhysRevLett.92.066802} {\bibfield  {journal}
  {\bibinfo  {journal} {Phys. Rev. Lett.}\ }\textbf {\bibinfo {volume} {92}},\
  \bibinfo {pages} {066802} (\bibinfo {year} {2004})}\BibitemShut {NoStop}%
\bibitem [{\citenamefont {Shaw}\ \emph {et~al.}(2008)\citenamefont {Shaw},
  \citenamefont {Lutchyn}, \citenamefont {Delsing},\ and\ \citenamefont
  {Echternach}}]{shaw2008kinetics}%
  \BibitemOpen
  \bibfield  {author} {\bibinfo {author} {\bibfnamefont {M.~D.}\ \bibnamefont
  {Shaw}}, \bibinfo {author} {\bibfnamefont {R.~M.}\ \bibnamefont {Lutchyn}},
  \bibinfo {author} {\bibfnamefont {P.}~\bibnamefont {Delsing}},\ and\ \bibinfo
  {author} {\bibfnamefont {P.~M.}\ \bibnamefont {Echternach}},\ }\bibfield
  {title} {\bibinfo {title} {Kinetics of nonequilibrium quasiparticle tunneling
  in superconducting charge qubits},\ }\href
  {https://doi.org/10.1103/PhysRevB.78.024503} {\bibfield  {journal} {\bibinfo
  {journal} {Phys. Rev. B}\ }\textbf {\bibinfo {volume} {78}},\ \bibinfo
  {pages} {024503} (\bibinfo {year} {2008})}\BibitemShut {NoStop}%
\bibitem [{\citenamefont {de~Visser}\ \emph {et~al.}(2011)\citenamefont
  {de~Visser}, \citenamefont {Baselmans}, \citenamefont {Diener}, \citenamefont
  {Yates}, \citenamefont {Endo},\ and\ \citenamefont
  {Klapwijk}}]{devisser2011number}%
  \BibitemOpen
  \bibfield  {author} {\bibinfo {author} {\bibfnamefont {P.~J.}\ \bibnamefont
  {de~Visser}}, \bibinfo {author} {\bibfnamefont {J.~J.~A.}\ \bibnamefont
  {Baselmans}}, \bibinfo {author} {\bibfnamefont {P.}~\bibnamefont {Diener}},
  \bibinfo {author} {\bibfnamefont {S.~J.~C.}\ \bibnamefont {Yates}}, \bibinfo
  {author} {\bibfnamefont {A.}~\bibnamefont {Endo}},\ and\ \bibinfo {author}
  {\bibfnamefont {T.~M.}\ \bibnamefont {Klapwijk}},\ }\bibfield  {title}
  {\bibinfo {title} {Number fluctuations of sparse quasiparticles in a
  superconductor},\ }\href {https://doi.org/10.1103/PhysRevLett.106.167004}
  {\bibfield  {journal} {\bibinfo  {journal} {Phys. Rev. Lett.}\ }\textbf
  {\bibinfo {volume} {106}},\ \bibinfo {pages} {167004} (\bibinfo {year}
  {2011})}\BibitemShut {NoStop}%
\bibitem [{\citenamefont {Martinis}\ \emph {et~al.}(2009)\citenamefont
  {Martinis}, \citenamefont {Ansmann},\ and\ \citenamefont
  {Aumentado}}]{martinis2009energy}%
  \BibitemOpen
  \bibfield  {author} {\bibinfo {author} {\bibfnamefont {J.~M.}\ \bibnamefont
  {Martinis}}, \bibinfo {author} {\bibfnamefont {M.}~\bibnamefont {Ansmann}},\
  and\ \bibinfo {author} {\bibfnamefont {J.}~\bibnamefont {Aumentado}},\
  }\bibfield  {title} {\bibinfo {title} {Energy decay in superconducting
  {Josephson}-junction qubits from nonequilibrium quasiparticle excitations},\
  }\href {https://doi.org/10.1103/PhysRevLett.103.097002} {\bibfield  {journal}
  {\bibinfo  {journal} {Phys. Rev. Lett.}\ }\textbf {\bibinfo {volume} {103}},\
  \bibinfo {pages} {097002} (\bibinfo {year} {2009})}\BibitemShut {NoStop}%
\bibitem [{\citenamefont {Catelani}\ \emph {et~al.}(2011)\citenamefont
  {Catelani}, \citenamefont {Koch}, \citenamefont {Frunzio}, \citenamefont
  {Schoelkopf}, \citenamefont {Devoret},\ and\ \citenamefont
  {Glazman}}]{catelani2011quasiparticle}%
  \BibitemOpen
  \bibfield  {author} {\bibinfo {author} {\bibfnamefont {G.}~\bibnamefont
  {Catelani}}, \bibinfo {author} {\bibfnamefont {J.}~\bibnamefont {Koch}},
  \bibinfo {author} {\bibfnamefont {L.}~\bibnamefont {Frunzio}}, \bibinfo
  {author} {\bibfnamefont {R.~J.}\ \bibnamefont {Schoelkopf}}, \bibinfo
  {author} {\bibfnamefont {M.~H.}\ \bibnamefont {Devoret}},\ and\ \bibinfo
  {author} {\bibfnamefont {L.~I.}\ \bibnamefont {Glazman}},\ }\bibfield
  {title} {\bibinfo {title} {Quasiparticle relaxation of superconducting qubits
  in the presence of flux},\ }\href
  {https://doi.org/10.1103/PhysRevLett.106.077002} {\bibfield  {journal}
  {\bibinfo  {journal} {Phys. Rev. Lett.}\ }\textbf {\bibinfo {volume} {106}},\
  \bibinfo {pages} {077002} (\bibinfo {year} {2011})}\BibitemShut {NoStop}%
\bibitem [{\citenamefont {Pop}\ \emph {et~al.}(2014)\citenamefont {Pop},
  \citenamefont {Geerlings}, \citenamefont {Catelani}, \citenamefont
  {Schoelkopf}, \citenamefont {Glazman},\ and\ \citenamefont
  {Devoret}}]{pop2014coherent}%
  \BibitemOpen
  \bibfield  {author} {\bibinfo {author} {\bibfnamefont {I.~M.}\ \bibnamefont
  {Pop}}, \bibinfo {author} {\bibfnamefont {K.}~\bibnamefont {Geerlings}},
  \bibinfo {author} {\bibfnamefont {G.}~\bibnamefont {Catelani}}, \bibinfo
  {author} {\bibfnamefont {R.~J.}\ \bibnamefont {Schoelkopf}}, \bibinfo
  {author} {\bibfnamefont {L.~I.}\ \bibnamefont {Glazman}},\ and\ \bibinfo
  {author} {\bibfnamefont {M.~H.}\ \bibnamefont {Devoret}},\ }\bibfield
  {title} {\bibinfo {title} {Coherent suppression of electromagnetic
  dissipation due to superconducting quasiparticles},\ }\href
  {https://doi.org/10.1038/nature13017} {\bibfield  {journal} {\bibinfo
  {journal} {Nature}\ }\textbf {\bibinfo {volume} {508}},\ \bibinfo {pages}
  {369} (\bibinfo {year} {2014})}\BibitemShut {NoStop}%
\bibitem [{\citenamefont {Wang}\ \emph {et~al.}(2014)\citenamefont {Wang},
  \citenamefont {Gao}, \citenamefont {Pop}, \citenamefont {Vool}, \citenamefont
  {Axline}, \citenamefont {Brecht}, \citenamefont {Heeres}, \citenamefont
  {Frunzio}, \citenamefont {Devoret}, \citenamefont {Catelani}, \citenamefont
  {Glazman},\ and\ \citenamefont {Schoelkopf}}]{wang2014measurement}%
  \BibitemOpen
  \bibfield  {author} {\bibinfo {author} {\bibfnamefont {C.}~\bibnamefont
  {Wang}}, \bibinfo {author} {\bibfnamefont {Y.~Y.}\ \bibnamefont {Gao}},
  \bibinfo {author} {\bibfnamefont {I.~M.}\ \bibnamefont {Pop}}, \bibinfo
  {author} {\bibfnamefont {U.}~\bibnamefont {Vool}}, \bibinfo {author}
  {\bibfnamefont {C.}~\bibnamefont {Axline}}, \bibinfo {author} {\bibfnamefont
  {T.}~\bibnamefont {Brecht}}, \bibinfo {author} {\bibfnamefont {R.~W.}\
  \bibnamefont {Heeres}}, \bibinfo {author} {\bibfnamefont {L.}~\bibnamefont
  {Frunzio}}, \bibinfo {author} {\bibfnamefont {M.~H.}\ \bibnamefont
  {Devoret}}, \bibinfo {author} {\bibfnamefont {G.}~\bibnamefont {Catelani}},
  \bibinfo {author} {\bibfnamefont {L.~I.}\ \bibnamefont {Glazman}},\ and\
  \bibinfo {author} {\bibfnamefont {R.~J.}\ \bibnamefont {Schoelkopf}},\
  }\bibfield  {title} {\bibinfo {title} {Measurement and control of
  quasiparticle dynamics in a superconducting qubit},\ }\href
  {https://doi.org/10.1038/ncomms6836} {\bibfield  {journal} {\bibinfo
  {journal} {Nature Communications}\ }\textbf {\bibinfo {volume} {5}},\
  \bibinfo {pages} {5836} (\bibinfo {year} {2014})}\BibitemShut {NoStop}%
\bibitem [{\citenamefont {Patel}\ \emph {et~al.}(2017)\citenamefont {Patel},
  \citenamefont {Pechenezhskiy}, \citenamefont {Plourde}, \citenamefont
  {Vavilov},\ and\ \citenamefont {McDermott}}]{patel2017phononmediated}%
  \BibitemOpen
  \bibfield  {author} {\bibinfo {author} {\bibfnamefont {U.}~\bibnamefont
  {Patel}}, \bibinfo {author} {\bibfnamefont {I.~V.}\ \bibnamefont
  {Pechenezhskiy}}, \bibinfo {author} {\bibfnamefont {B.~L.~T.}\ \bibnamefont
  {Plourde}}, \bibinfo {author} {\bibfnamefont {M.~G.}\ \bibnamefont
  {Vavilov}},\ and\ \bibinfo {author} {\bibfnamefont {R.}~\bibnamefont
  {McDermott}},\ }\bibfield  {title} {\bibinfo {title} {Phonon-mediated
  quasiparticle poisoning of superconducting microwave resonators},\ }\href
  {https://doi.org/10.1103/PhysRevB.96.220501} {\bibfield  {journal} {\bibinfo
  {journal} {Phys. Rev. B}\ }\textbf {\bibinfo {volume} {96}},\ \bibinfo
  {pages} {220501(R)} (\bibinfo {year} {2017})}\BibitemShut {NoStop}%
\bibitem [{\citenamefont {Gustavsson}\ \emph {et~al.}(2016)\citenamefont
  {Gustavsson}, \citenamefont {Yan}, \citenamefont {Catelani}, \citenamefont
  {Bylander}, \citenamefont {Kamal}, \citenamefont {Birenbaum}, \citenamefont
  {Hover}, \citenamefont {Rosenberg}, \citenamefont {Samach}, \citenamefont
  {Sears}, \citenamefont {Weber}, \citenamefont {Yoder}, \citenamefont
  {Clarke}, \citenamefont {Kerman}, \citenamefont {Yoshihara}, \citenamefont
  {Nakamura}, \citenamefont {Orlando},\ and\ \citenamefont
  {Oliver}}]{gustavsson2016suppressing}%
  \BibitemOpen
  \bibfield  {author} {\bibinfo {author} {\bibfnamefont {S.}~\bibnamefont
  {Gustavsson}}, \bibinfo {author} {\bibfnamefont {F.}~\bibnamefont {Yan}},
  \bibinfo {author} {\bibfnamefont {G.}~\bibnamefont {Catelani}}, \bibinfo
  {author} {\bibfnamefont {J.}~\bibnamefont {Bylander}}, \bibinfo {author}
  {\bibfnamefont {A.}~\bibnamefont {Kamal}}, \bibinfo {author} {\bibfnamefont
  {J.}~\bibnamefont {Birenbaum}}, \bibinfo {author} {\bibfnamefont
  {D.}~\bibnamefont {Hover}}, \bibinfo {author} {\bibfnamefont
  {D.}~\bibnamefont {Rosenberg}}, \bibinfo {author} {\bibfnamefont
  {G.}~\bibnamefont {Samach}}, \bibinfo {author} {\bibfnamefont {A.~P.}\
  \bibnamefont {Sears}}, \bibinfo {author} {\bibfnamefont {S.~J.}\ \bibnamefont
  {Weber}}, \bibinfo {author} {\bibfnamefont {J.~L.}\ \bibnamefont {Yoder}},
  \bibinfo {author} {\bibfnamefont {J.}~\bibnamefont {Clarke}}, \bibinfo
  {author} {\bibfnamefont {A.~J.}\ \bibnamefont {Kerman}}, \bibinfo {author}
  {\bibfnamefont {F.}~\bibnamefont {Yoshihara}}, \bibinfo {author}
  {\bibfnamefont {Y.}~\bibnamefont {Nakamura}}, \bibinfo {author}
  {\bibfnamefont {T.~P.}\ \bibnamefont {Orlando}},\ and\ \bibinfo {author}
  {\bibfnamefont {W.~D.}\ \bibnamefont {Oliver}},\ }\bibfield  {title}
  {\bibinfo {title} {Suppressing relaxation in superconducting qubits by
  quasiparticle pumping},\ }\href {https://doi.org/10.1126/science.aah5844}
  {\bibfield  {journal} {\bibinfo  {journal} {Science}\ }\textbf {\bibinfo
  {volume} {354}},\ \bibinfo {pages} {1573} (\bibinfo {year}
  {2016})}\BibitemShut {NoStop}%
\bibitem [{\citenamefont {Ferguson}(2008)}]{ferguson2008quasiparticle}%
  \BibitemOpen
  \bibfield  {author} {\bibinfo {author} {\bibfnamefont {A.~J.}\ \bibnamefont
  {Ferguson}},\ }\bibfield  {title} {\bibinfo {title} {Quasiparticle cooling of
  a single {Cooper} pair transistor},\ }\href
  {https://doi.org/10.1063/1.2968214} {\bibfield  {journal} {\bibinfo
  {journal} {Applied Physics Letters}\ }\textbf {\bibinfo {volume} {93}},\
  \bibinfo {pages} {052501} (\bibinfo {year} {2008})}\BibitemShut {NoStop}%
\bibitem [{\citenamefont {Higginbotham}\ \emph {et~al.}(2015)\citenamefont
  {Higginbotham}, \citenamefont {Albrecht}, \citenamefont {Kir{\v{s}}anskas},
  \citenamefont {Chang}, \citenamefont {Kuemmeth}, \citenamefont {Krogstrup},
  \citenamefont {Jespersen}, \citenamefont {Nyg{\aa}rd}, \citenamefont
  {Flensberg},\ and\ \citenamefont {Marcus}}]{higginbotham2015parity}%
  \BibitemOpen
  \bibfield  {author} {\bibinfo {author} {\bibfnamefont {A.~P.}\ \bibnamefont
  {Higginbotham}}, \bibinfo {author} {\bibfnamefont {S.~M.}\ \bibnamefont
  {Albrecht}}, \bibinfo {author} {\bibfnamefont {G.}~\bibnamefont
  {Kir{\v{s}}anskas}}, \bibinfo {author} {\bibfnamefont {W.}~\bibnamefont
  {Chang}}, \bibinfo {author} {\bibfnamefont {F.}~\bibnamefont {Kuemmeth}},
  \bibinfo {author} {\bibfnamefont {P.}~\bibnamefont {Krogstrup}}, \bibinfo
  {author} {\bibfnamefont {T.~S.}\ \bibnamefont {Jespersen}}, \bibinfo {author}
  {\bibfnamefont {J.}~\bibnamefont {Nyg{\aa}rd}}, \bibinfo {author}
  {\bibfnamefont {K.}~\bibnamefont {Flensberg}},\ and\ \bibinfo {author}
  {\bibfnamefont {C.~M.}\ \bibnamefont {Marcus}},\ }\bibfield  {title}
  {\bibinfo {title} {Parity lifetime of bound states in a proximitized
  semiconductor nanowire},\ }\href {https://doi.org/10.1038/nphys3461}
  {\bibfield  {journal} {\bibinfo  {journal} {Nature Physics}\ }\textbf
  {\bibinfo {volume} {11}},\ \bibinfo {pages} {1017} (\bibinfo {year}
  {2015})}\BibitemShut {NoStop}%
\bibitem [{\citenamefont {Vool}\ \emph {et~al.}(2014)\citenamefont {Vool},
  \citenamefont {Pop}, \citenamefont {Sliwa}, \citenamefont {Abdo},
  \citenamefont {Wang}, \citenamefont {Brecht}, \citenamefont {Gao},
  \citenamefont {Shankar}, \citenamefont {Hatridge}, \citenamefont {Catelani},
  \citenamefont {Mirrahimi}, \citenamefont {Frunzio}, \citenamefont
  {Schoelkopf}, \citenamefont {Glazman},\ and\ \citenamefont
  {Devoret}}]{vool2014nonpoissonian}%
  \BibitemOpen
  \bibfield  {author} {\bibinfo {author} {\bibfnamefont {U.}~\bibnamefont
  {Vool}}, \bibinfo {author} {\bibfnamefont {I.~M.}\ \bibnamefont {Pop}},
  \bibinfo {author} {\bibfnamefont {K.}~\bibnamefont {Sliwa}}, \bibinfo
  {author} {\bibfnamefont {B.}~\bibnamefont {Abdo}}, \bibinfo {author}
  {\bibfnamefont {C.}~\bibnamefont {Wang}}, \bibinfo {author} {\bibfnamefont
  {T.}~\bibnamefont {Brecht}}, \bibinfo {author} {\bibfnamefont {Y.~Y.}\
  \bibnamefont {Gao}}, \bibinfo {author} {\bibfnamefont {S.}~\bibnamefont
  {Shankar}}, \bibinfo {author} {\bibfnamefont {M.}~\bibnamefont {Hatridge}},
  \bibinfo {author} {\bibfnamefont {G.}~\bibnamefont {Catelani}}, \bibinfo
  {author} {\bibfnamefont {M.}~\bibnamefont {Mirrahimi}}, \bibinfo {author}
  {\bibfnamefont {L.}~\bibnamefont {Frunzio}}, \bibinfo {author} {\bibfnamefont
  {R.~J.}\ \bibnamefont {Schoelkopf}}, \bibinfo {author} {\bibfnamefont
  {L.~I.}\ \bibnamefont {Glazman}},\ and\ \bibinfo {author} {\bibfnamefont
  {M.~H.}\ \bibnamefont {Devoret}},\ }\bibfield  {title} {\bibinfo {title}
  {Non-{Poissonian} quantum jumps of a fluxonium qubit due to quasiparticle
  excitations},\ }\href {https://doi.org/10.1103/PhysRevLett.113.247001}
  {\bibfield  {journal} {\bibinfo  {journal} {Phys. Rev. Lett.}\ }\textbf
  {\bibinfo {volume} {113}},\ \bibinfo {pages} {247001} (\bibinfo {year}
  {2014})}\BibitemShut {NoStop}%
\bibitem [{\citenamefont {Veps{\"a}l{\"a}inen}\ \emph
  {et~al.}(2020)\citenamefont {Veps{\"a}l{\"a}inen}, \citenamefont {Karamlou},
  \citenamefont {Orrell}, \citenamefont {Dogra}, \citenamefont {Loer},
  \citenamefont {Vasconcelos}, \citenamefont {Kim}, \citenamefont {Melville},
  \citenamefont {Niedzielski}, \citenamefont {Yoder}, \citenamefont
  {Gustavsson}, \citenamefont {Formaggio}, \citenamefont {VanDevender},\ and\
  \citenamefont {Oliver}}]{vepsalainen2020impact}%
  \BibitemOpen
  \bibfield  {author} {\bibinfo {author} {\bibfnamefont {A.}~\bibnamefont
  {Veps{\"a}l{\"a}inen}}, \bibinfo {author} {\bibfnamefont {A.~H.}\
  \bibnamefont {Karamlou}}, \bibinfo {author} {\bibfnamefont {J.~L.}\
  \bibnamefont {Orrell}}, \bibinfo {author} {\bibfnamefont {A.~S.}\
  \bibnamefont {Dogra}}, \bibinfo {author} {\bibfnamefont {B.}~\bibnamefont
  {Loer}}, \bibinfo {author} {\bibfnamefont {F.}~\bibnamefont {Vasconcelos}},
  \bibinfo {author} {\bibfnamefont {D.~K.}\ \bibnamefont {Kim}}, \bibinfo
  {author} {\bibfnamefont {A.~J.}\ \bibnamefont {Melville}}, \bibinfo {author}
  {\bibfnamefont {B.~M.}\ \bibnamefont {Niedzielski}}, \bibinfo {author}
  {\bibfnamefont {J.~L.}\ \bibnamefont {Yoder}}, \bibinfo {author}
  {\bibfnamefont {S.}~\bibnamefont {Gustavsson}}, \bibinfo {author}
  {\bibfnamefont {J.~A.}\ \bibnamefont {Formaggio}}, \bibinfo {author}
  {\bibfnamefont {B.~A.}\ \bibnamefont {VanDevender}},\ and\ \bibinfo {author}
  {\bibfnamefont {W.~D.}\ \bibnamefont {Oliver}},\ }\bibfield  {title}
  {\bibinfo {title} {Impact of ionizing radiation on superconducting qubit
  coherence},\ }\href {https://doi.org/10.1038/s41586-020-2619-8} {\bibfield
  {journal} {\bibinfo  {journal} {Nature}\ }\textbf {\bibinfo {volume} {584}},\
  \bibinfo {pages} {551} (\bibinfo {year} {2020})}\BibitemShut {NoStop}%
\bibitem [{\citenamefont {Cardani}\ \emph {et~al.}(2021)\citenamefont
  {Cardani}, \citenamefont {Valenti}, \citenamefont {Casali}, \citenamefont
  {Catelani}, \citenamefont {Charpentier}, \citenamefont {Clemenza},
  \citenamefont {Colantoni}, \citenamefont {Cruciani}, \citenamefont {Gironi},
  \citenamefont {Gr{\"u}nhaupt}, \citenamefont {Gusenkova}, \citenamefont
  {Henriques}, \citenamefont {Lagoin}, \citenamefont {Martinez}, \citenamefont
  {Pettinari}, \citenamefont {Rusconi}, \citenamefont {Sander}, \citenamefont
  {Ustinov}, \citenamefont {Weber}, \citenamefont {Wernsdorfer}, \citenamefont
  {Vignati}, \citenamefont {Pirro},\ and\ \citenamefont
  {Pop}}]{cardani2020reducing}%
  \BibitemOpen
  \bibfield  {author} {\bibinfo {author} {\bibfnamefont {L.}~\bibnamefont
  {Cardani}}, \bibinfo {author} {\bibfnamefont {F.}~\bibnamefont {Valenti}},
  \bibinfo {author} {\bibfnamefont {N.}~\bibnamefont {Casali}}, \bibinfo
  {author} {\bibfnamefont {G.}~\bibnamefont {Catelani}}, \bibinfo {author}
  {\bibfnamefont {T.}~\bibnamefont {Charpentier}}, \bibinfo {author}
  {\bibfnamefont {M.}~\bibnamefont {Clemenza}}, \bibinfo {author}
  {\bibfnamefont {I.}~\bibnamefont {Colantoni}}, \bibinfo {author}
  {\bibfnamefont {A.}~\bibnamefont {Cruciani}}, \bibinfo {author}
  {\bibfnamefont {L.}~\bibnamefont {Gironi}}, \bibinfo {author} {\bibfnamefont
  {L.}~\bibnamefont {Gr{\"u}nhaupt}}, \bibinfo {author} {\bibfnamefont
  {D.}~\bibnamefont {Gusenkova}}, \bibinfo {author} {\bibfnamefont
  {F.}~\bibnamefont {Henriques}}, \bibinfo {author} {\bibfnamefont
  {M.}~\bibnamefont {Lagoin}}, \bibinfo {author} {\bibfnamefont
  {M.}~\bibnamefont {Martinez}}, \bibinfo {author} {\bibfnamefont
  {G.}~\bibnamefont {Pettinari}}, \bibinfo {author} {\bibfnamefont
  {C.}~\bibnamefont {Rusconi}}, \bibinfo {author} {\bibfnamefont
  {O.}~\bibnamefont {Sander}}, \bibinfo {author} {\bibfnamefont {A.~V.}\
  \bibnamefont {Ustinov}}, \bibinfo {author} {\bibfnamefont {M.}~\bibnamefont
  {Weber}}, \bibinfo {author} {\bibfnamefont {W.}~\bibnamefont {Wernsdorfer}},
  \bibinfo {author} {\bibfnamefont {M.}~\bibnamefont {Vignati}}, \bibinfo
  {author} {\bibfnamefont {S.}~\bibnamefont {Pirro}},\ and\ \bibinfo {author}
  {\bibfnamefont {I.~M.}\ \bibnamefont {Pop}},\ }\bibfield  {title} {\bibinfo
  {title} {Reducing the impact of radioactivity on quantum circuits in a
  deep-underground facility},\ }\href
  {https://doi.org/10.1038/s41467-021-23032-z} {\bibfield  {journal} {\bibinfo
  {journal} {Nature Communications}\ }\textbf {\bibinfo {volume} {12}},\
  \bibinfo {pages} {2733} (\bibinfo {year} {2021})}\BibitemShut {NoStop}%
\bibitem [{\citenamefont {Wilen}\ \emph {et~al.}(2021)\citenamefont {Wilen},
  \citenamefont {Abdullah}, \citenamefont {Kurinsky}, \citenamefont {Stanford},
  \citenamefont {Cardani}, \citenamefont {D'Imperio}, \citenamefont {Tomei},
  \citenamefont {Faoro}, \citenamefont {Ioffe}, \citenamefont {Liu},
  \citenamefont {Opremcak}, \citenamefont {Christensen}, \citenamefont
  {DuBois},\ and\ \citenamefont {McDermott}}]{wilen2020correlated}%
  \BibitemOpen
  \bibfield  {author} {\bibinfo {author} {\bibfnamefont {C.~D.}\ \bibnamefont
  {Wilen}}, \bibinfo {author} {\bibfnamefont {S.}~\bibnamefont {Abdullah}},
  \bibinfo {author} {\bibfnamefont {N.~A.}\ \bibnamefont {Kurinsky}}, \bibinfo
  {author} {\bibfnamefont {C.}~\bibnamefont {Stanford}}, \bibinfo {author}
  {\bibfnamefont {L.}~\bibnamefont {Cardani}}, \bibinfo {author} {\bibfnamefont
  {G.}~\bibnamefont {D'Imperio}}, \bibinfo {author} {\bibfnamefont
  {C.}~\bibnamefont {Tomei}}, \bibinfo {author} {\bibfnamefont
  {L.}~\bibnamefont {Faoro}}, \bibinfo {author} {\bibfnamefont {L.~B.}\
  \bibnamefont {Ioffe}}, \bibinfo {author} {\bibfnamefont {C.~H.}\ \bibnamefont
  {Liu}}, \bibinfo {author} {\bibfnamefont {A.}~\bibnamefont {Opremcak}},
  \bibinfo {author} {\bibfnamefont {B.~G.}\ \bibnamefont {Christensen}},
  \bibinfo {author} {\bibfnamefont {J.~L.}\ \bibnamefont {DuBois}},\ and\
  \bibinfo {author} {\bibfnamefont {R.}~\bibnamefont {McDermott}},\ }\bibfield
  {title} {\bibinfo {title} {Correlated charge noise and relaxation errors in
  superconducting qubits},\ }\href {https://doi.org/10.1038/s41586-021-03557-5}
  {\bibfield  {journal} {\bibinfo  {journal} {Nature}\ }\textbf {\bibinfo
  {volume} {594}},\ \bibinfo {pages} {369} (\bibinfo {year}
  {2021})}\BibitemShut {NoStop}%
\bibitem [{\citenamefont {Martinis}(2021)}]{martinis2020saving}%
  \BibitemOpen
  \bibfield  {author} {\bibinfo {author} {\bibfnamefont {J.~M.}\ \bibnamefont
  {Martinis}},\ }\bibfield  {title} {\bibinfo {title} {Saving superconducting
  quantum processors from qubit decay and correlated errors generated by gamma
  and cosmic rays},\ }\href {https://doi.org/10.1038/s41534-021-00431-0}
  {\bibfield  {journal} {\bibinfo  {journal} {npj Quantum Information}\
  }\textbf {\bibinfo {volume} {7}},\ \bibinfo {pages} {90} (\bibinfo {year}
  {2021})}\BibitemShut {NoStop}%
\bibitem [{\citenamefont {Wilson}\ \emph {et~al.}(2001)\citenamefont {Wilson},
  \citenamefont {Frunzio},\ and\ \citenamefont
  {Prober}}]{wilson2001timeresolved}%
  \BibitemOpen
  \bibfield  {author} {\bibinfo {author} {\bibfnamefont {C.~M.}\ \bibnamefont
  {Wilson}}, \bibinfo {author} {\bibfnamefont {L.}~\bibnamefont {Frunzio}},\
  and\ \bibinfo {author} {\bibfnamefont {D.~E.}\ \bibnamefont {Prober}},\
  }\bibfield  {title} {\bibinfo {title} {Time-resolved measurements of
  thermodynamic fluctuations of the particle number in a nondegenerate {Fermi}
  gas},\ }\href {https://doi.org/10.1103/PhysRevLett.87.067004} {\bibfield
  {journal} {\bibinfo  {journal} {Phys. Rev. Lett.}\ }\textbf {\bibinfo
  {volume} {87}},\ \bibinfo {pages} {067004} (\bibinfo {year}
  {2001})}\BibitemShut {NoStop}%
\bibitem [{\citenamefont {Lambert}\ \emph {et~al.}(2014)\citenamefont
  {Lambert}, \citenamefont {Edwards}, \citenamefont {Esmail}, \citenamefont
  {Pollock}, \citenamefont {Barrett}, \citenamefont {Lovett},\ and\
  \citenamefont {Ferguson}}]{lambert2014experimental}%
  \BibitemOpen
  \bibfield  {author} {\bibinfo {author} {\bibfnamefont {N.~J.}\ \bibnamefont
  {Lambert}}, \bibinfo {author} {\bibfnamefont {M.}~\bibnamefont {Edwards}},
  \bibinfo {author} {\bibfnamefont {A.~A.}\ \bibnamefont {Esmail}}, \bibinfo
  {author} {\bibfnamefont {F.~A.}\ \bibnamefont {Pollock}}, \bibinfo {author}
  {\bibfnamefont {S.~D.}\ \bibnamefont {Barrett}}, \bibinfo {author}
  {\bibfnamefont {B.~W.}\ \bibnamefont {Lovett}},\ and\ \bibinfo {author}
  {\bibfnamefont {A.~J.}\ \bibnamefont {Ferguson}},\ }\bibfield  {title}
  {\bibinfo {title} {Experimental observation of the breaking and recombination
  of single {Cooper} pairs},\ }\href
  {https://doi.org/10.1103/PhysRevB.90.140503} {\bibfield  {journal} {\bibinfo
  {journal} {Phys. Rev. B}\ }\textbf {\bibinfo {volume} {90}},\ \bibinfo
  {pages} {140503} (\bibinfo {year} {2014})}\BibitemShut {NoStop}%
\bibitem [{\citenamefont {Serniak}\ \emph {et~al.}(2019)\citenamefont
  {Serniak}, \citenamefont {Diamond}, \citenamefont {Hays}, \citenamefont
  {Fatemi}, \citenamefont {Shankar}, \citenamefont {Frunzio}, \citenamefont
  {Schoelkopf},\ and\ \citenamefont {Devoret}}]{serniak2019direct}%
  \BibitemOpen
  \bibfield  {author} {\bibinfo {author} {\bibfnamefont {K.}~\bibnamefont
  {Serniak}}, \bibinfo {author} {\bibfnamefont {S.}~\bibnamefont {Diamond}},
  \bibinfo {author} {\bibfnamefont {M.}~\bibnamefont {Hays}}, \bibinfo {author}
  {\bibfnamefont {V.}~\bibnamefont {Fatemi}}, \bibinfo {author} {\bibfnamefont
  {S.}~\bibnamefont {Shankar}}, \bibinfo {author} {\bibfnamefont
  {L.}~\bibnamefont {Frunzio}}, \bibinfo {author} {\bibfnamefont
  {R.}~\bibnamefont {Schoelkopf}},\ and\ \bibinfo {author} {\bibfnamefont
  {M.}~\bibnamefont {Devoret}},\ }\bibfield  {title} {\bibinfo {title} {Direct
  dispersive monitoring of charge parity in offset-charge-sensitive
  transmons},\ }\href {https://doi.org/10.1103/PhysRevApplied.12.014052}
  {\bibfield  {journal} {\bibinfo  {journal} {Phys. Rev. Applied}\ }\textbf
  {\bibinfo {volume} {12}},\ \bibinfo {pages} {014052} (\bibinfo {year}
  {2019})}\BibitemShut {NoStop}%
\bibitem [{\citenamefont {Van~Woerkom}\ \emph {et~al.}(2015)\citenamefont
  {Van~Woerkom}, \citenamefont {Geresdi},\ and\ \citenamefont
  {Kouwenhoven}}]{vanwoerkom2015one}%
  \BibitemOpen
  \bibfield  {author} {\bibinfo {author} {\bibfnamefont {D.~J.}\ \bibnamefont
  {Van~Woerkom}}, \bibinfo {author} {\bibfnamefont {A.}~\bibnamefont
  {Geresdi}},\ and\ \bibinfo {author} {\bibfnamefont {L.~P.}\ \bibnamefont
  {Kouwenhoven}},\ }\bibfield  {title} {\bibinfo {title} {One minute parity
  lifetime of a {NbTiN} {Cooper}-pair transistor},\ }\href
  {https://doi.org/10.1038/nphys3342} {\bibfield  {journal} {\bibinfo
  {journal} {Nature Physics}\ }\textbf {\bibinfo {volume} {11}},\ \bibinfo
  {pages} {547} (\bibinfo {year} {2015})}\BibitemShut {NoStop}%
\bibitem [{\citenamefont {Hays}\ \emph {et~al.}(2018)\citenamefont {Hays},
  \citenamefont {de~Lange}, \citenamefont {Serniak}, \citenamefont {van
  Woerkom}, \citenamefont {Bouman}, \citenamefont {Krogstrup}, \citenamefont
  {Nyg\aa{}rd}, \citenamefont {Geresdi},\ and\ \citenamefont
  {Devoret}}]{hays2018direct}%
  \BibitemOpen
  \bibfield  {author} {\bibinfo {author} {\bibfnamefont {M.}~\bibnamefont
  {Hays}}, \bibinfo {author} {\bibfnamefont {G.}~\bibnamefont {de~Lange}},
  \bibinfo {author} {\bibfnamefont {K.}~\bibnamefont {Serniak}}, \bibinfo
  {author} {\bibfnamefont {D.~J.}\ \bibnamefont {van Woerkom}}, \bibinfo
  {author} {\bibfnamefont {D.}~\bibnamefont {Bouman}}, \bibinfo {author}
  {\bibfnamefont {P.}~\bibnamefont {Krogstrup}}, \bibinfo {author}
  {\bibfnamefont {J.}~\bibnamefont {Nyg\aa{}rd}}, \bibinfo {author}
  {\bibfnamefont {A.}~\bibnamefont {Geresdi}},\ and\ \bibinfo {author}
  {\bibfnamefont {M.~H.}\ \bibnamefont {Devoret}},\ }\bibfield  {title}
  {\bibinfo {title} {Direct microwave measurement of {Andreev}-bound-state
  dynamics in a semiconductor-nanowire {Josephson} junction},\ }\href
  {https://doi.org/10.1103/PhysRevLett.121.047001} {\bibfield  {journal}
  {\bibinfo  {journal} {Phys. Rev. Lett.}\ }\textbf {\bibinfo {volume} {121}},\
  \bibinfo {pages} {047001} (\bibinfo {year} {2018})}\BibitemShut {NoStop}%
\bibitem [{\citenamefont {Karzig}\ \emph {et~al.}(2021)\citenamefont {Karzig},
  \citenamefont {Cole},\ and\ \citenamefont
  {Pikulin}}]{karzig2020quasiparticle}%
  \BibitemOpen
  \bibfield  {author} {\bibinfo {author} {\bibfnamefont {T.}~\bibnamefont
  {Karzig}}, \bibinfo {author} {\bibfnamefont {W.~S.}\ \bibnamefont {Cole}},\
  and\ \bibinfo {author} {\bibfnamefont {D.~I.}\ \bibnamefont {Pikulin}},\
  }\bibfield  {title} {\bibinfo {title} {Quasiparticle poisoning of {Majorana}
  qubits},\ }\href {https://doi.org/10.1103/PhysRevLett.126.057702} {\bibfield
  {journal} {\bibinfo  {journal} {Phys. Rev. Lett.}\ }\textbf {\bibinfo
  {volume} {126}},\ \bibinfo {pages} {057702} (\bibinfo {year}
  {2021})}\BibitemShut {NoStop}%
\bibitem [{\citenamefont {Barends}\ \emph {et~al.}(2011)\citenamefont
  {Barends}, \citenamefont {Wenner}, \citenamefont {Lenander}, \citenamefont
  {Chen}, \citenamefont {Bialczak}, \citenamefont {Kelly}, \citenamefont
  {Lucero}, \citenamefont {O’Malley}, \citenamefont {Mariantoni},
  \citenamefont {Sank}, \citenamefont {Wang}, \citenamefont {White},
  \citenamefont {Yin}, \citenamefont {Zhao}, \citenamefont {Cleland},
  \citenamefont {Martinis},\ and\ \citenamefont
  {Baselmans}}]{barends2011minimizing}%
  \BibitemOpen
  \bibfield  {author} {\bibinfo {author} {\bibfnamefont {R.}~\bibnamefont
  {Barends}}, \bibinfo {author} {\bibfnamefont {J.}~\bibnamefont {Wenner}},
  \bibinfo {author} {\bibfnamefont {M.}~\bibnamefont {Lenander}}, \bibinfo
  {author} {\bibfnamefont {Y.}~\bibnamefont {Chen}}, \bibinfo {author}
  {\bibfnamefont {R.~C.}\ \bibnamefont {Bialczak}}, \bibinfo {author}
  {\bibfnamefont {J.}~\bibnamefont {Kelly}}, \bibinfo {author} {\bibfnamefont
  {E.}~\bibnamefont {Lucero}}, \bibinfo {author} {\bibfnamefont
  {P.}~\bibnamefont {O’Malley}}, \bibinfo {author} {\bibfnamefont
  {M.}~\bibnamefont {Mariantoni}}, \bibinfo {author} {\bibfnamefont
  {D.}~\bibnamefont {Sank}}, \bibinfo {author} {\bibfnamefont {H.}~\bibnamefont
  {Wang}}, \bibinfo {author} {\bibfnamefont {T.~C.}\ \bibnamefont {White}},
  \bibinfo {author} {\bibfnamefont {Y.}~\bibnamefont {Yin}}, \bibinfo {author}
  {\bibfnamefont {J.}~\bibnamefont {Zhao}}, \bibinfo {author} {\bibfnamefont
  {A.~N.}\ \bibnamefont {Cleland}}, \bibinfo {author} {\bibfnamefont {J.~M.}\
  \bibnamefont {Martinis}},\ and\ \bibinfo {author} {\bibfnamefont {J.~J.~A.}\
  \bibnamefont {Baselmans}},\ }\bibfield  {title} {\bibinfo {title} {Minimizing
  quasiparticle generation from stray infrared light in superconducting quantum
  circuits},\ }\href {https://doi.org/10.1063/1.3638063} {\bibfield  {journal}
  {\bibinfo  {journal} {Appl. Phys. Lett.}\ }\textbf {\bibinfo {volume} {99}},\
  \bibinfo {pages} {113507} (\bibinfo {year} {2011})}\BibitemShut {NoStop}%
\bibitem [{\citenamefont {Saira}\ \emph {et~al.}(2012)\citenamefont {Saira},
  \citenamefont {Kemppinen}, \citenamefont {Maisi},\ and\ \citenamefont
  {Pekola}}]{saira2012vanishing}%
  \BibitemOpen
  \bibfield  {author} {\bibinfo {author} {\bibfnamefont {O.-P.}\ \bibnamefont
  {Saira}}, \bibinfo {author} {\bibfnamefont {A.}~\bibnamefont {Kemppinen}},
  \bibinfo {author} {\bibfnamefont {V.~F.}\ \bibnamefont {Maisi}},\ and\
  \bibinfo {author} {\bibfnamefont {J.~P.}\ \bibnamefont {Pekola}},\ }\bibfield
   {title} {\bibinfo {title} {Vanishing quasiparticle density in a hybrid
  {Al/Cu/Al} single-electron transistor},\ }\href
  {https://doi.org/10.1103/PhysRevB.85.012504} {\bibfield  {journal} {\bibinfo
  {journal} {Phys. Rev. B}\ }\textbf {\bibinfo {volume} {85}},\ \bibinfo
  {pages} {012504} (\bibinfo {year} {2012})}\BibitemShut {NoStop}%
\bibitem [{\citenamefont {Maisi}\ \emph {et~al.}(2013)\citenamefont {Maisi},
  \citenamefont {Lotkhov}, \citenamefont {Kemppinen}, \citenamefont {Heimes},
  \citenamefont {Muhonen},\ and\ \citenamefont {Pekola}}]{maisi2013excitation}%
  \BibitemOpen
  \bibfield  {author} {\bibinfo {author} {\bibfnamefont {V.~F.}\ \bibnamefont
  {Maisi}}, \bibinfo {author} {\bibfnamefont {S.~V.}\ \bibnamefont {Lotkhov}},
  \bibinfo {author} {\bibfnamefont {A.}~\bibnamefont {Kemppinen}}, \bibinfo
  {author} {\bibfnamefont {A.}~\bibnamefont {Heimes}}, \bibinfo {author}
  {\bibfnamefont {J.~T.}\ \bibnamefont {Muhonen}},\ and\ \bibinfo {author}
  {\bibfnamefont {J.~P.}\ \bibnamefont {Pekola}},\ }\bibfield  {title}
  {\bibinfo {title} {Excitation of single quasiparticles in a small
  superconducting {Al} island connected to normal-metal leads by tunnel
  junctions},\ }\href {https://doi.org/10.1103/PhysRevLett.111.147001}
  {\bibfield  {journal} {\bibinfo  {journal} {Phys. Rev. Lett.}\ }\textbf
  {\bibinfo {volume} {111}},\ \bibinfo {pages} {147001} (\bibinfo {year}
  {2013})}\BibitemShut {NoStop}%
\bibitem [{\citenamefont {Pobell}(2007)}]{pobell}%
  \BibitemOpen
  \bibfield  {author} {\bibinfo {author} {\bibfnamefont {F.}~\bibnamefont
  {Pobell}},\ }\href@noop {} {\emph {\bibinfo {title} {Matter and Methods at
  Low Temperatures}}},\ \bibinfo {edition} {3rd}\ ed.\ (\bibinfo  {publisher}
  {Springer-Verlag},\ \bibinfo {year} {2007})\BibitemShut {NoStop}%
\bibitem [{\citenamefont {Pekola}\ \emph {et~al.}(2010)\citenamefont {Pekola},
  \citenamefont {Maisi}, \citenamefont {Kafanov}, \citenamefont {Chekurov},
  \citenamefont {Kemppinen}, \citenamefont {Pashkin}, \citenamefont {Saira},
  \citenamefont {M\"ott\"onen},\ and\ \citenamefont
  {Tsai}}]{pekola2010environment}%
  \BibitemOpen
  \bibfield  {author} {\bibinfo {author} {\bibfnamefont {J.~P.}\ \bibnamefont
  {Pekola}}, \bibinfo {author} {\bibfnamefont {V.~F.}\ \bibnamefont {Maisi}},
  \bibinfo {author} {\bibfnamefont {S.}~\bibnamefont {Kafanov}}, \bibinfo
  {author} {\bibfnamefont {N.}~\bibnamefont {Chekurov}}, \bibinfo {author}
  {\bibfnamefont {A.}~\bibnamefont {Kemppinen}}, \bibinfo {author}
  {\bibfnamefont {Y.~A.}\ \bibnamefont {Pashkin}}, \bibinfo {author}
  {\bibfnamefont {O.-P.}\ \bibnamefont {Saira}}, \bibinfo {author}
  {\bibfnamefont {M.}~\bibnamefont {M\"ott\"onen}},\ and\ \bibinfo {author}
  {\bibfnamefont {J.~S.}\ \bibnamefont {Tsai}},\ }\bibfield  {title} {\bibinfo
  {title} {Environment-assisted tunneling as an origin of the {Dynes} density
  of states},\ }\href {https://doi.org/10.1103/PhysRevLett.105.026803}
  {\bibfield  {journal} {\bibinfo  {journal} {Phys. Rev. Lett.}\ }\textbf
  {\bibinfo {volume} {105}},\ \bibinfo {pages} {026803} (\bibinfo {year}
  {2010})}\BibitemShut {NoStop}%
\bibitem [{\citenamefont {Koski}\ \emph {et~al.}(2011)\citenamefont {Koski},
  \citenamefont {Peltonen}, \citenamefont {Meschke},\ and\ \citenamefont
  {Pekola}}]{koski2011laterally}%
  \BibitemOpen
  \bibfield  {author} {\bibinfo {author} {\bibfnamefont {J.~V.}\ \bibnamefont
  {Koski}}, \bibinfo {author} {\bibfnamefont {J.~T.}\ \bibnamefont {Peltonen}},
  \bibinfo {author} {\bibfnamefont {M.}~\bibnamefont {Meschke}},\ and\ \bibinfo
  {author} {\bibfnamefont {J.~P.}\ \bibnamefont {Pekola}},\ }\bibfield  {title}
  {\bibinfo {title} {Laterally proximized aluminum tunnel junctions},\ }\href
  {https://doi.org/10.1063/1.3590922} {\bibfield  {journal} {\bibinfo
  {journal} {Applied Physics Letters}\ }\textbf {\bibinfo {volume} {98}},\
  \bibinfo {pages} {203501} (\bibinfo {year} {2011})}\BibitemShut {NoStop}%
\bibitem [{\citenamefont {Zorin}(1995)}]{zorin1995thermocoax}%
  \BibitemOpen
  \bibfield  {author} {\bibinfo {author} {\bibfnamefont {A.~B.}\ \bibnamefont
  {Zorin}},\ }\bibfield  {title} {\bibinfo {title} {The thermocoax cable as the
  microwave frequency filter for single electron circuits},\ }\href
  {https://doi.org/10.1063/1.1145385} {\bibfield  {journal} {\bibinfo
  {journal} {Review of Scientific Instruments}\ }\textbf {\bibinfo {volume}
  {66}},\ \bibinfo {pages} {4296} (\bibinfo {year} {1995})}\BibitemShut
  {NoStop}%
\bibitem [{\citenamefont {Schoelkopf}\ \emph {et~al.}(1998)\citenamefont
  {Schoelkopf}, \citenamefont {Wahlgren}, \citenamefont {Kozhevnikov},
  \citenamefont {Delsing},\ and\ \citenamefont
  {Prober}}]{schoelkopf1998radiofrequency}%
  \BibitemOpen
  \bibfield  {author} {\bibinfo {author} {\bibfnamefont {R.~J.}\ \bibnamefont
  {Schoelkopf}}, \bibinfo {author} {\bibfnamefont {P.}~\bibnamefont
  {Wahlgren}}, \bibinfo {author} {\bibfnamefont {A.~A.}\ \bibnamefont
  {Kozhevnikov}}, \bibinfo {author} {\bibfnamefont {P.}~\bibnamefont
  {Delsing}},\ and\ \bibinfo {author} {\bibfnamefont {D.~E.}\ \bibnamefont
  {Prober}},\ }\bibfield  {title} {\bibinfo {title} {The radio-frequency
  single-electron transistor {(RF-SET)}: A fast and ultrasensitive
  electrometer},\ }\href {https://doi.org/10.1126/science.280.5367.1238}
  {\bibfield  {journal} {\bibinfo  {journal} {Science}\ }\textbf {\bibinfo
  {volume} {280}},\ \bibinfo {pages} {1238} (\bibinfo {year}
  {1998})}\BibitemShut {NoStop}%
\bibitem [{\citenamefont {Simbierowicz}\ \emph {et~al.}(2018)\citenamefont
  {Simbierowicz}, \citenamefont {Vesterinen}, \citenamefont {Gr{\"o}nberg},
  \citenamefont {Lehtinen}, \citenamefont {Prunnila},\ and\ \citenamefont
  {Hassel}}]{simbierowicz2018fluxdriven}%
  \BibitemOpen
  \bibfield  {author} {\bibinfo {author} {\bibfnamefont {S.}~\bibnamefont
  {Simbierowicz}}, \bibinfo {author} {\bibfnamefont {V.}~\bibnamefont
  {Vesterinen}}, \bibinfo {author} {\bibfnamefont {L.}~\bibnamefont
  {Gr{\"o}nberg}}, \bibinfo {author} {\bibfnamefont {J.}~\bibnamefont
  {Lehtinen}}, \bibinfo {author} {\bibfnamefont {M.}~\bibnamefont {Prunnila}},\
  and\ \bibinfo {author} {\bibfnamefont {J.}~\bibnamefont {Hassel}},\
  }\bibfield  {title} {\bibinfo {title} {A flux-driven {Josephson} parametric
  amplifier for {sub-GHz} frequencies fabricated with side-wall passivated
  spacer junction technology},\ }\href
  {https://doi.org/10.1088/1361-6668/aad4f2} {\bibfield  {journal} {\bibinfo
  {journal} {Superconductor Science and Technology}\ }\textbf {\bibinfo
  {volume} {31}},\ \bibinfo {pages} {105001} (\bibinfo {year}
  {2018})}\BibitemShut {NoStop}%
\end{thebibliography}


\end{document}

\title{A superconductor free of quasiparticles for seconds \\ Supplementary Material}

\author{E. T. Mannila}
\email[]{elsa.mannila@aalto.fi}
\altaffiliation{Present address: VTT Technical Research Centre of Finland Ltd, QTF Centre of Excellence, P.O. Box 1000, FI-02044 VTT, Finland}
\affiliation{QTF Centre of Excellence, Department of Applied Physics, Aalto University, FI-00076 Aalto, Finland}

\author{P. Samuelsson}
\affiliation{Physics Department and NanoLund, Lund University, Box 118, 22100 Lund, Sweden}

\author{S. Simbierowicz}
\altaffiliation{Present address: Bluefors Oy, Arinatie 10, 00370 Helsinki, Finland}
\affiliation{VTT Technical Research Centre of Finland Ltd, QTF Centre of Excellence, P.O. Box 1000, FI-02044 VTT, Finland}

\author{J. T. Peltonen}
\affiliation{QTF Centre of Excellence, Department of Applied Physics, Aalto University, FI-00076 Aalto, Finland}

\author{V. Vesterinen}
\affiliation{VTT Technical Research Centre of Finland Ltd, QTF Centre of Excellence, P.O. Box 1000, FI-02044 VTT, Finland}

\author{L. Gr{\"o}nberg}
\affiliation{VTT Technical Research Centre of Finland Ltd, QTF Centre of Excellence, P.O. Box 1000, FI-02044 VTT, Finland}

\author{J. Hassel}
\altaffiliation{Present address: IQM, Espoo, Finland}
\affiliation{VTT Technical Research Centre of Finland Ltd, QTF Centre of Excellence, P.O. Box 1000, FI-02044 VTT, Finland}

\author{V. F. Maisi}
\affiliation{Physics Department and NanoLund, Lund University, Box 118, 22100 Lund, Sweden}

\author{J. P. Pekola}
\affiliation{QTF Centre of Excellence, Department of Applied Physics, Aalto University, FI-00076 Aalto, Finland}

\date{\today}
\maketitle

\section{Supplementary Note 1: Sample fabrication}

\begin{figure}
\includegraphics{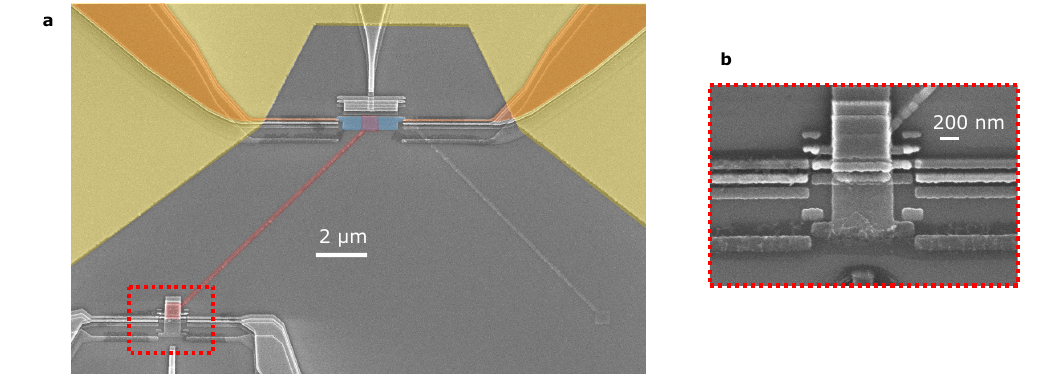}
\caption{\textbf{a} False-color scanning electron micrograph of a device lithographically identical to the measured device. The ground planes (2 nm Ti/30 nm Au/2 nm Ti, colored yellow), and the capacitive coupler electrode (10 nm Cr, colored red) are covered by a 40 nm thick insulating aluminum oxide layer grown by atomic layer deposition. The other coupler electrode visible on the right side of the sample is not used in this experiment. 
\textbf{b} Close up of charge detector. 
\label{sfig:sem}}
\end{figure}

We start the sample fabrication by depositing Ti/Au/Ti ground planes (yellow in Supplementary Fig. \ref{sfig:sem}\textbf{a}) and alignment markers 2 nm/30 nm/2 nm thick on a 525 $\mu$m thick silicon substrate covered by 300 nm of thermal silicon oxide. In a second lithography step, we deposit a 10 nm thick and 10 $\mu$m long chromium wire (diagonal red line in Fig. \ref{sfig:sem}\textbf{a}), which couples the superconducting island and charge detector capacitively. %
The resistance of the chromium wire is estimated to be tens of k$\Omega$ based on DC measurements of similar wires, whose resistance increased roughly 10\% at low temperatures compared to the room-temperature value. 
The layers are covered by an insulating layer of 40 nm thick aluminum oxide grown by atomic layer deposition. 

The superconducting island and charge detector were fabricated simultaneously in a third electron-beam lithography step using a Ge-based hard mask and four-angle evaporation. 
We evaporate first 35 nm of aluminum, forming the superconducting island. The tunnel junctions are formed by oxidizing in 2 mbar of O$_2$ for 90 s, followed by deposition of 35 nm of copper which forms the normal metal leads of the superconducting island. Next, we evaporate a further 20 nm of aluminum, oxidize for 30 s in 1 mbar of O2 and evaporate 60 nm of copper to form the charge detector. We use the laterally proximized junction technique \cite{koski2011laterally} to create a normal-state charge detector with aluminum oxide tunnel junctions: the 20 nm thick aluminum film is inversely proximized by the clean contact to the 35 nm copper film. We expect this choice to minimize detector backaction due to nonequilibrium phonons \cite{patel2017phononmediated}. Using chromium as the material of the capacitive coupler electrode was motivated by avoiding backaction, as the resistive coupler is expected to filter the high-frequency components of shot noise that could break Cooper pairs \cite{saira2012vanishing}. However, we do not have quantitative evidence that the coupler material reduces backaction.

The design of the sample is such that the superconducting island does not overlap with any of its shadow copies produced by the multi-angle evaporation, and its normal leads overlap with their copies only at a distance 1.5 $\mu$m away, far enough to suppress any potential proximity effect. 

% Coverage: 0.2 mm2 from ground planes, another 0.2 mm2 from detectors x 12 devices on the chip
We note that the normal metal layers of the device, as well as the gold ground plane, may serve as phonon traps \cite{henriques2019phonon, karatsu2019mitigation}. These layers cover approximately 40\% of the chip surface within the mm-scale immediate vicinity of the device and roughly 20\% of the entire chip with an area of roughly 1 cm$^2$. The chip is attached to the copper sample holder with vacuum grease, and is in direct contact with the sample holder through its entire backside. We thus expect that a large fraction of high-energy phonons, expected to be created by cosmic rays or ionizing radiation hitting the substrate, will either escape into the sample holder or be absorbed in the normal metal traps, rather than creating quasiparticles on the superconducting island.

It is possible that the time-dependent process breaking Cooper pairs would originate from the sample itself, for instance from the relaxation of two-level systems or perhaps strain on the island. The fabrication methods and the feature size of the aluminum island are similar to those commonly used in superconducting qubits, although it is also possible that the ortho-para conversion in the copper leads could contribute to the time-dependent decay.

\clearpage
\section{Supplementary Note 2: Measurement setup}

\begin{figure}
\includegraphics{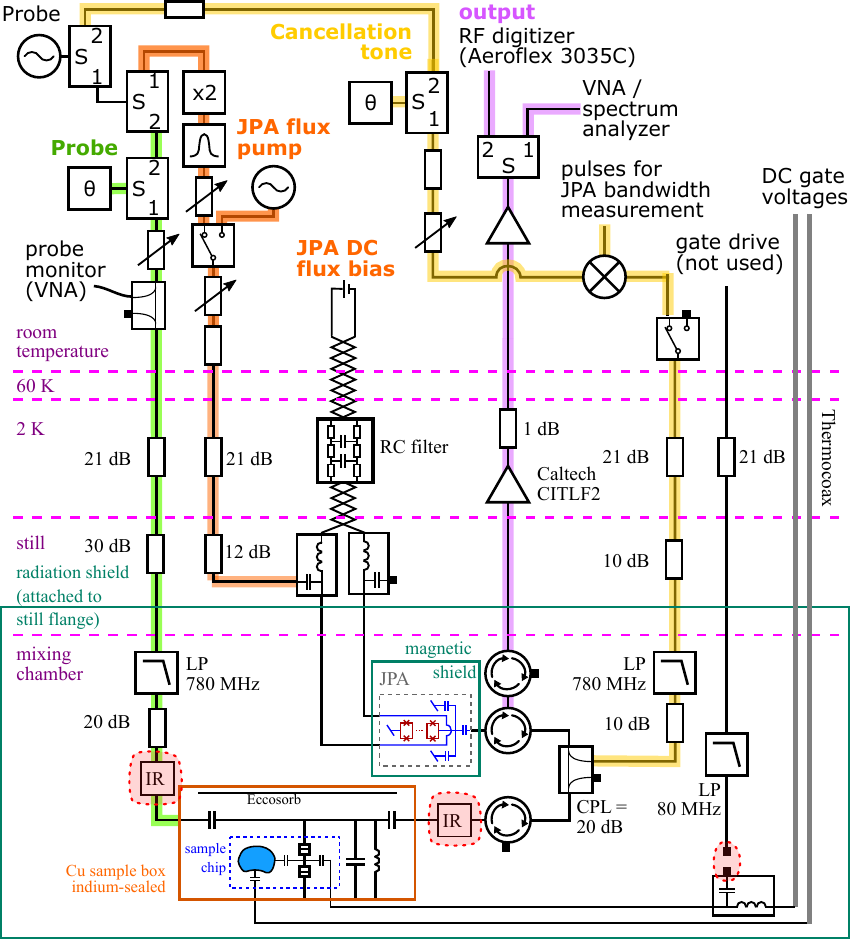}
\caption{\label{sfig:meas-schematic}
	Schematic of the measurement circuitry, filtering and shielding used in the experiment. The components highlighted with pink circles were added to the setup for the second cooldown (see text). 
}
\end{figure}

Figure \ref{sfig:meas-schematic} shows a schematic of the measurement setup used. As described in Methods, the DC lines used to tune the gates of the superconducting island and the charge detector are filtered with 1 to 2 m of Thermocoax \cite{zorin1995thermocoax}. The RF input lines (probe tone (highlighted in green), cancellation tone (yellow), JPA pump (orange)) are attenuated at the different temperature stages, and circulators provide approximately 40 dB isolation between the sample and JPA. 
The probe and cancellation are low-pass filtered at the mixing chamber stage with commercial filters (Mini-Circuits VLFX), while for the second cooldown we added home-made Eccosorb filters closest to the sample holder (boxes labeled "IR"). The same sample holder, refrigerator and a RF setup with similar components have been used in electron thermometry experiments where thermalization has been verified down to below 30 mK \cite{viisanen2015incomplete, karimi2018noninvasive}, while in this experiment we were able to verify the thermalization of the normal metal leads down to 100 mK (see Fig. \ref{sfig:gate-dependence}). For the second cooldown, we also disconnected the RF drive line, not used in the present experiments, at the RF input of the bias tee at the mixing chamber stage and terminated the port with a 50~$\Omega$ cryotermination (indicated in pink). 

As shown in Fig. 4 of the main text, in both cooldowns the main source of quasiparticles was a time-dependent process not affected by these changes in the setup. Insufficient filtering of radiation from higher-temperature stages of the cryostat would be expected to lead to an elevated Cooper pair breaking rate, but without the time dependence. We find imperfect thermalization of the components a possible but unlikely explanation for the decay, as the electron temperature of the normal metal leads was constant over time. The time-dependent contribution could originate from, for instance, ortho-para conversion of hydrogen molecules in the copper sample stage, or radioactive contaminants re-introduced into the system by a thermal cycle. Other possible sources related to the measurement setup are two-level systems in the various amorphous materials in the sample holder and its vicinity.

We use a room-temperature setup similar to that used in Ref. \cite{simbierowicz2018fluxdriven} to create the probe, interferometric cancellation and JPA pump tones from a single RF generator. An additional pump signal generator visible in Fig. \ref{sfig:meas-schematic} was used for initial JPA tuning in the phase-preserving mode, but it was switched off for the measurements presented here. 
The output signal (purple) is amplified at 2 K and at room temperature and measured with either a vector network analyser (initial detector characterization), spectrum analyser (sensitivity measurements) or RF digitizer (time-domain traces).

\section{Supplementary Note 3: Charge detector and Josephson parametric amplifier characterization}

\subsection{Charge detector}

\begin{figure}
\includegraphics{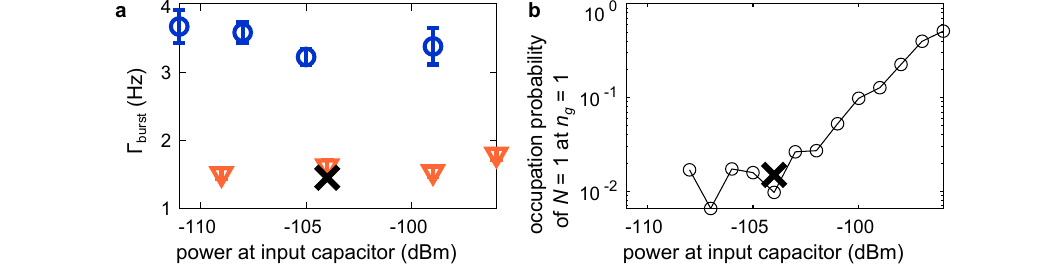}
\caption{\label{sfig:backaction}
\textbf{a} Burst rates at $n_g=0$ as a function of the probe power applied, measured 54 (circles) or 123 (triangles) days after the start of the first cooldown. The cross indicates the data in Figs. 1-3 of the main text. Error bars reflect statistical uncertainty due to the number of bursts measured at each point.
\textbf{b} At higher input powers than used in the main text, the backaction increases the population of the charge state $N=1$ at $n_g=1$, although the burst rate at $n_g =0$ is unaffected (see Fig. \ref{sfig:gate-dependence} for $2e$-periodic occupation probabilities of the charge states versus $n_g$).
}
\end{figure}

The charge detector, shown in Fig. \ref{sfig:sem}\textbf{b}, is a normal-metallic single-electron transistor with room-temperature resistance 55~k$\Omega$. It is connected in parallel to a resonant circuit on a separate silicon chip, formed of an evaporated aluminum spiral inductor and mostly parasitic capacitance, with the input and output signal coupled through finger capacitors, similarly to the RF thermometer described in Ref. \cite{viisanen2015incomplete} and the RF-SET used in Ref. \cite{mannila2019detecting}. The resonant frequency of the circuit was 580.35 MHz in the first cooldown and  581.28 MHz in the second. We attribute this change to slight shifting of the chips or bond wires and corresponding changes in parasitic capacitance due to the thermal cycle. 
We apply a probe signal at the resonant frequency $f_{probe}$ and adjust the probe power and gate offset of the detector for maximum sensitivity, but do not apply a DC voltage bias. The probe power used in the data presented in the main text was -104 dBm at the input capacitor (Fig. \ref{sfig:backaction}). 

\subsection{JPA operation}

The output of the charge detector is amplified by the Josephson parametric amplifier, which is otherwise similar to the devices presented in Ref. \cite{simbierowicz2018fluxdriven} but has a larger gain-bandwidth product. 
We operate the amplifier in the phase-sensitive mode, where the pump frequency $f_{pump} = 2f_{probe}$, and tune to between 18 and 30 dB gain in the amplified quadrature (18 dB for the data shown in Figs. 1-3). 
We cancel the steady-state probe tone interferometrically to overcome the dynamic range limitations of the JPA, which manifest as excess noise apparent in the time domain signal even at powers smaller than the 1-dB saturation point. The cancellation tone is combined with the probe tone by a directional coupler between the sample and JPA, as shown in Fig. \ref{sfig:meas-schematic}. 
In practice, we first fix the pump signal and DC flux bias to the JPA as well as the magnitude of the probe signal. We then tune the phase of the probe signal with the cancellation tone off such that the charge response is in the amplified quadrature and adjust the gate offset  of the charge detector to a sensitive operating point. 
Then, we adjust the phase and amplitude of the cancellation tone with the JPA pump off such that the output amplitude corresponding to the $N=0$ charge state is close to zero.

\subsection{Charge sensitivity and signal-to-noise ratio of charge readout}

\begin{figure}
\includegraphics[width=\textwidth]{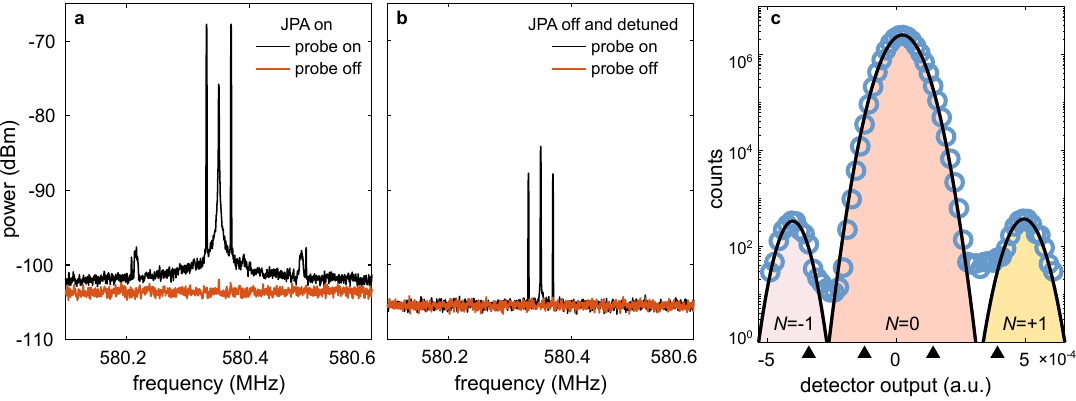}
\caption{\label{sfig:sensitivity}
Sensitivity measurement with JPA on \textbf{a} and with JPA off and detuned \textbf{b}. \textbf{c} Histogram of the signal levels of the filtered trace shown in Fig. \ref{sfig:trace-analysis} (circles), while solid lines are Gaussian fits corresponding to the three charge states. The arrows indicate the thresholds used for determining the charge states (see Supplementary Note 4).
}
\end{figure}

In Fig. \ref{sfig:sensitivity}(a,b), we show the response of our charge detector to an $0.04e$ RMS excitation at 20 kHz applied to the detector through the DC gate, measured with a spectrum analyser bandwidth of 1 kHz. 
The line labeled "RF drive" in Fig. \ref{sfig:meas-schematic} was not used in this experiment due to uncertainty of its transmission at low frequencies. Here, the JPA gain is 20.5 dB and the input probe tone to the sample is -101~dBm (compared to 18 dB and -104 dBm for the data shown in the main text). The signal-to-noise ratios are 31 dB and 18 dB with the JPA on and off, corresponding to sensitivities of $2.3 \times 10^{-5}$ $e/\sqrt{\text{Hz}}$ and $1.1 \times 10^{-4}$ $e/\sqrt{\text{Hz}}$, respectively. For the JPA off measurement, we have also detuned the JPA flux bias so that the JPA resonance is moved outside the frequency range of interest. The broad noise peak visible around the carrier tone arises from the noise of the detector but also from true tunneling events, as the detector is coupled to the superconducting island during this measurement as well.
The peaks in the spectrum at 130 kHz from the carrier arise from electronic pickup that couples to the gate of the superconducting island which we were unable to eliminate in the experiment. 
For comparison, we also show the measured spectra when the probe and cancellation tones are turned off. With the JPA off, the noise floor is dominated by the amplifier (Caltech CITLF2) at the 2 K stage of the cryostat, which has a noise temperature of 3.5 K. 
Although the measured charge sensitivities are about two orders of magnitude worse than the state of the art \cite{schaal2020fast}, they are sufficient for single-shot detection of charge jumps of roughly $0.1e$, corresponding to single-electron tunneling events in the superconducting island, with high fidelity. 

The operating point of the detector and JPA in the charge sensitivity measurement shown above is not exactly the same as the one used in the data shown in Figs. 1-3 of the main text, and these results are merely indicative of our general detector performance. We additionally characterize the signal-to-noise of the charge measurement directly from the traces measured in the time domain. In Fig. \ref{sfig:sensitivity}\textbf{c}, we show a histogram of a filtered time trace from the dataset used in the main text, with Gaussian fits corresponding to the three charge states observed. The finite count of points between the separate peaks originate from the transitions between the peaks, where the signal spends time in intermediate values due to the filtering. Defining the signal-to-noise ratio like in Ref. \cite{razmadze2019radiofrequency} as the ratio of the separation of the peaks to the standard deviation of the fitted Gaussian distributions, we obtain signal-to-noise ratios of 6.5 and 6.7 for the transitions $N=-1\leftrightarrow 0 $ and  $N=+1\leftrightarrow 0 $, respectively. In this experiment, we need a high signal-to-noise ratio to be able to distinguish reliably noise from the odd charge states, which are occupied with a probability on the order of $10^{-4}$. As the Gaussian fits in Fig. \ref{sfig:sensitivity}\textbf{c} do not overlap, the probability of interpreting experimental noise as spurious transitions is small. 

\subsection{Detection bandwidth}

The 3-dB bandwidth of the resonant circuit is approximately 1 MHz. We measure the bandwidth of the JPA in the phase-sensitive mode similarly as in the supplementary material of Ref. \cite{riste2012initialization} by applying square pulses to the cancellation line with similar amplitude as the charge detection signal, while the probe signal is turned off. Even with 30 dB gain, the JPA rise time is less than a microsecond. 
The resistive chromium wire used to provide the capacitive coupling between the charge detector and the island (see Fig. \ref{sfig:sem}) is not expected to limit the bandwidth of the charge detector, as the cutoff frequency of the resulting $RC$ filter is estimated to be on the order of 1 GHz.  
Under these conditions, the detector bandwidth is limited by the post-processing low-pass filter, as shown in Fig. \ref{sfig:trace-analysis} below.

\clearpage
\section{Supplementary Note 4: Data analysis}

\subsection{Filtering}
For the data shown in Figs. 1-3 of the main text, we record  5 s traces of the detector output with a sampling rate of 4 MHz and filter the signal digitally to achieve a high signal-to-noise ratio. For some of the data shown in the supplementary material and Fig. 4, we have used also slightly longer or shorter traces and sampling rates between 2 MHz and 10 MHz. For the data shown in Figs. 1-3 of the main text, we first subtract a moving median of the signal over 10 ms to remove low-frequency drifts. Then, we apply a digital low-pass filter with a cutoff of 150 kHz, which sets the detector rise time to $3~\mu$s. This low-pass filter sets the effective bandwidth of the charge detector, and this filtered data is the detector output shown in Fig. 1 of the main text. We do not use the moving median filter in cases where more than one charge state is occupied with a probability larger than 1\%, such as in obtaining the $n_g$ dependence of the tunneling rates shown below. 

\subsection{Determining the instantaneous charge state}
\begin{figure}
    \centering
    \includegraphics{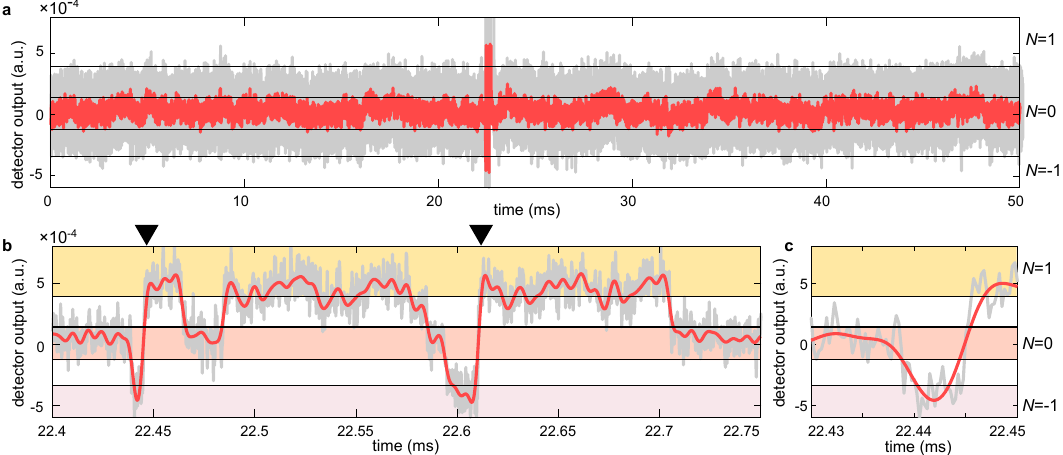}
    \caption{\textbf{a} Portion of the timetrace histogrammed in Fig. \ref{sfig:sensitivity}, zoomed in \textbf{b} to cover a single burst of tunneling and to show individual tunnel events more clearly in \textbf{c}. The filtered trace is shown in red, while the light gray trace is the rotated IQ trace before low-pass filtering. The shaded areas and horizontal lines indicate the thresholds for the different charge states, which are also indicated as arrows in Fig. \ref{sfig:sensitivity}\textbf{c}. In panel \textbf{b}, the triangles indicate two-electron Andreev tunneling events.
    }
    \label{sfig:trace-analysis}
\end{figure}

We rotate the measured IQ signal such that the charge response is only along one quadrature and analyse the traces one by one. To identify the signal levels corresponding to each charge state, we filter the signal with a cutoff frequency of 50 kHz, histogram the signal values in this filtered trace and identify peaks in the histogram with a minimum separation of $3 \times 10^{-4}$ in the units on the x-axis of Fig. \ref{sfig:sensitivity}\textbf{c}. We exclude the traces where the algorithm has not correctly identified 3 peaks corresponding to the three populated charge states at $n_g = 0$. We then define threshold signal values halfway between the peaks and the minima between each adjacent pair of peaks in the histogram, which are shown in Figs. \ref{sfig:sensitivity}\textbf{c} and \ref{sfig:trace-analysis}. Then, we use a hysteretic trigger to assign the charge states based on the trace filtered with a cutoff frequency of 150 kHz: a transition from charge state $N$ to $N'$ is considered to have happened only when the signal value crosses the threshold corresponding to state $N'$. %This is illustrated in Fig. \ref{sfig:trace-analysis}. 

For the data acquired during the second cooldown, shown in Fig. 4 in the main text, we additionally identify the times when quasiparticle trapping or de-trapping events occur, as described in Supplementary Note 8 below. 

\subsection{Identifying tunneling events and bursts}
After identifying the charge states from the raw data, we identify the tunneling events as single-electron or two-electron tunneling events. We classify a tunneling event as Andreev tunneling if the signal changes from $N=+1$ to $-1$ or vice versa within the detector rise time 3~$\mu$s, ascertained from the measured waiting time distribution in the state $N=0$ similarly as in Ref. \cite{maisi2011realtime}. The Andreev tunneling rates in the data shown in Figs. 1-3 in the main text are $\Gamma_{-1\rightarrow +1} = 5.3$~kHz and $\Gamma_{+1\rightarrow -1} = 0.6$~kHz. On average, less than one Andreev event occurs per burst, and the Andreev events do not affect our conclusions. Even if we would include these events in the analysis as single-electron events, the distribution of number of 1e tunneling events per burst (Fig. 2\textbf{b} of main text) would still appear exponential, although the best fit parameter changes to approximately 0.35, and we would still observe an initial decay in the tunneling rates within the bursts followed by saturation to finite values (Fig. 3\textbf{b} of the main text). 

We split the time traces into bursts and intervals between bursts with the condition that tunneling events separated by at least $t_{\text{sep}} = 250~\mu$s in charge state $N=0$ belong to separate bursts. With this choice, to interpret two successive bursts erraneously as one requires that a second Cooper-pair breaking event happens within time $t_{\text{burst}} + t_{\text{sep}}$ from the start of the burst, where $t_{\text{burst}} = 170~\mu$s is the mean length of a burst. Given that the Cooper pair breaking events occur according to a Poisson process with rate $\Gamma_{\text{burst}} = 1.5$~Hz, this probability is $1-\exp [-\Gamma_\text{burst} (t_{\text{burst}} + t_{\text{sep}})] = 6 \times 10^{-4}$.
On the other hand, the probability of splitting a long burst into two is given by the probability that the system were in state $N=0$ and $N_{\text{QP}} \geq 2$ for time $t_{\text{sep}}$ without a tunneling event occurring. The tunneling rate out of the state $N=0$ and $N_\text{QP}=2$ is $4\Gamma_{\text{QP}} = 32$~kHz, so this probability is $\exp (-4 \Gamma_{\text{QP}} t_{\text{sep}}) = 3 \times 10^{-4}$. Hence the total probability of misinterpreting bursts is on the order of $10^{-3}$. After identifying the bursts and tunneling events, Figs. 2\textbf{b} and 3\textbf{a} of the main text are directly obtained from counting the number of tunneling events. 

In plotting the waiting times between the bursts of tunneling in Fig. 2\textbf{a} of the main text, we leave out the intervals at the beginning and end of each 5 s long time trace. Due to the finite length of the time traces, long quiet times are underrepresented in the measured data, which means the fit for the rate parameter is an upper limit.  In Fig. 4, we plot instead the burst rate calculated as dividing the number of bursts by the total measurement time, which produces an unbiased estimate of the burst rate. The error bars in Figs. 2 and 4 represent the statistical uncertainty arising from the counts of events in each bin.

\subsection{Effective tunneling rates within the bursts}

\begin{figure}
    \centering
    \includegraphics{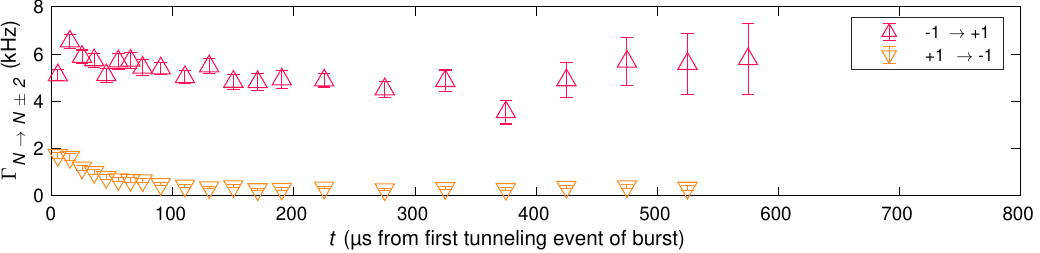}
    \caption{Andreev tunneling rates within the bursts of tunneling from the same dataset as the single-electron rates shown in Fig. 3\textbf{b} of the main text. Error bars reflect statistical uncertainty due to the number of tunneling events per time bin, and data is shown only if there are at least 10 bursts with events in the corresponding bin.}
    \label{sfig:andreev-within}
\end{figure}

We calculate effective tunneling rates within bursts $\Gamma_{N\rightarrow N'}(t)$ from charge states $N$ to $N'$ as a function time $t$ from the first tunneling event of a burst. For single-electron transitions, $N'=N \pm 1$, for Andreev rates, $N' = N \pm 2$. 
The rates are calculated as averages over time bins by dividing the number of transitions from $N$ to $N'$ within a bin by the total time spent in state $N$ within that bin. The time spent in state $N=0$ after a burst has ended is not included in the denominator, so that the rates are calculated only within the bursts. The bin widths are adjusted from $10~\mu$s at short times to $50~\mu$s at $t>250~\mu$s, as there are less long bursts available for calculating statistics. We show only those data points where we have statistics calculated from at least 10 bursts with events in the corresponding bin. The error bars in Fig. 3\textbf{b} of the main text and Figs. \ref{sfig:andreev-within} and \ref{sfig:gate-dependence} indicate statistical uncertainty based the count of events in each bin. 

The single-electron tunneling rates are shown in Fig. 3\textbf{b} of the main text, while the Andreev tunneling rates are shown in Fig. \ref{sfig:andreev-within}. The rate $\Gamma_{-1 \rightarrow +1}(t)$ is constant within the burst. This is expected theoretically, as the Andreev tunneling events are insensitive to the number of quasiparticles present and thus should be constant within the bursts. The slight decrease of the rate $\Gamma_{+1 \rightarrow -1}(t)$ at $t<100~\mu$s is due to misinterpretation of single-electron transitions as Andreev, and the effect is larger because there are less true Andreev events in this direction. In Fig. \ref{sfig:backtunneling}, discussed below, we show the measured occupation probabilities of the charge states calculated as averages over the same time bins as the tunneling rates, again not including the time after a burst has ended.

Note that in both panels of Fig. 3 of the main text as well as Figs. \ref{sfig:andreev-within} and \ref{sfig:gate-dependence}, time $t$ is measured from the first tunneling event of each burst. This occurs typically within less than $30~\mu$s after the initial Cooper-pair-breaking event, as the rate out of the state with $N_{\text{QP}}=2$ is $4\Gamma_{\text{QP}} = 32$~kHz.

\subsection{Effect of finite bandwidth on the measurement}

Based on the detector performance above, we can estimate the probability that a burst of tunneling would be undetected due to finite bandwidth of the charge detector \cite{naaman2006poisson}. The shortest burst is one which starts with only two quasiparticles and correspondingly has only two tunneling events ($N=0 \rightarrow \pm 1 \rightarrow 0$), where the second event is expected to happen with the single-quasiparticle tunneling rate $\Gamma_{\text{QP}}=8$~kHz. Such an event will be missed if the system stays in the state $N = \pm 1$ for less than the detector response time of 3 $\mu$s, with probability $1-\exp (3~\mu\text{s}\times 8~\text{kHz}) \approx 2\%$. Incorporating the results of the modeling that takes into account the distribution of number of Cooper pairs broken, we estimate that we miss completely only 1.4\% of the bursts. Hence we can reliably identify the quasiparticle-free periods from the quiet periods in the time traces. 

After a burst of tunneling is correctly identified by the analysis procedure, it is possible that some of the tunneling events within the burst are missed. This is especially likely in the beginnings of the burst, where many quasiparticles are present and the tunneling rates are faster, meaning that the number of tunneling events tends to be underestimated for bursts where many Cooper pairs were initially broken. In particular, we can estimate that if there are more than 20 quasiparticles initially created, the decay rate by tunneling becomes faster than our detector bandwidth for the beginning of the burst. Since we observe essentially no counts between 10 to 20 quasiparticles nor excessive amount of cases at the tail of the exponential distribution if Fig. 2\textbf{b}, we however conclude that the cases with more than 20 quasiparticles are absent. Such high number of quasiparticles could be generated by high-energy phonons created by high-energy impacts. As discussed in the "Sample fabrication" section above, our sample structure is such that the high-energy phonons are expected to escape to the sample holder or the normal metal traps rather than creating quasiparticles on our superconducting island, which is in line with the absence of the large quasiparticle number cases.

In Fig. 2\textbf{b} of the main text, we also see that there are less bursts with $N_{\text{CP}} \geq 6$ than the exponential fit predicts. We believe that this may be caused by the finite detection bandwidth, even if the underlying distribution of Cooper pairs broken is exponential, as we are more likely to miss some tunneling events in the fast initial relaxation if many quasiparticles were initially created. 
To quantify this effect, we have generated simulated time traces from the model assuming an exponential distribution of broken Cooper pairs $N_{\text{CP}}$ and analysed this simulated data with the same procedure as the experimental data, which qualitatively reproduces the effect. Hence also our estimate for the underlying parameter $\lambda = 0.9$ characterizing the quasiparticle creation process (see Fig. 2\textbf{b} of main text) is an upper limit. Similarly, the proportion 40\% of events that break more than one Cooper pair is a lower limit. The simulations also confirm that the dip in the effective tunneling rates $\Gamma_{0\rightarrow \pm1}$ visible in Fig. 3\textbf{b} at $t<10~\mu$s is due to the finite detector bandwidth.

%We can estimate that the initial tunneling rates are higher than our detector bandwidth for $N_{\text{QP}}=10$ or more broken Cooper pairs. This means that we cannot directly count 

% which leads to the apparent dip in the rates $\Gamma_{0\rightarrow \pm 1}$ at $t<10~\mu$s in Fig. 3\textbf{b} of the main text.

\clearpage
\section{Supplementary Note 5: Rate equation model}

\begin{figure}[h]
	\begin{center}
		\centering \includegraphics[width=0.6\textwidth]{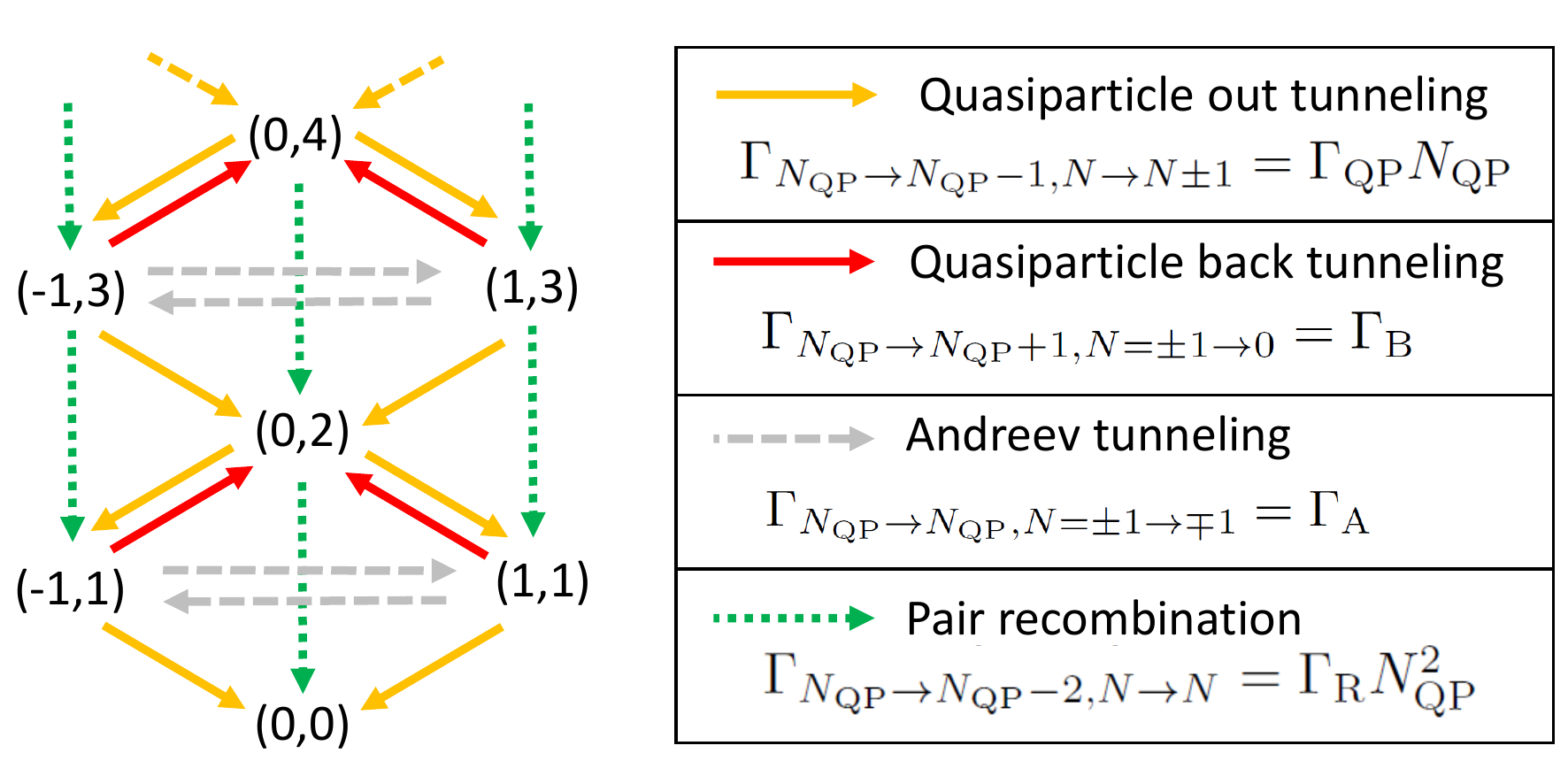}
		\caption{Left:  Charge-quasiparticle states $(N,N_{\text{QP}})$ with processes for transfer between states denoted with arrows. Right: List of state transfer processes with corresponding rates.}
		\label{sfig:Methodtheoryfig}
	\end{center}
\end{figure}

We describe the combined quasiparticle and charge dynamics within a burst with a rate equation for the joint probability $P(N_{\text{QP}},N,t)$ to have $N_{\text{QP}}=0,1,2,3...$ quasiparticles and $N=-1,0,1$ excess electron charge on the superconductor, at a given time $t$. Charge states with $|N| \geq 2$ have a neglible occupation probability due to strong Coulomb blockade, see discussion below. 
The rate equation, presented in Ref. \cite{maisi2013excitation}, is given by
\begin{equation}
\frac{dP(N_{\text{QP}},N,t)}{dt}=\sum_{N'_{\text{QP}},N'}\Gamma_{N'_{\text{QP}} \rightarrow N_{\text{QP}},N' \rightarrow N}P(N'_{\text{QP}},N',t) 
- \sum_{N'_{\text{QP}},N'}\Gamma_{N_{\text{QP}} \rightarrow N'_{\text{QP}},N \rightarrow N'}P(N_{\text{QP}},N,t).
\label{eq:rateeq}
\end{equation} 
The charge-quasiparticle states, denoted  $(N,N_{\text{QP}})$, and the four different transfer processes $N_{\text{QP}} \rightarrow N_{\text{QP}}', N  \rightarrow N'$ accounted for, with corresponding rates, are shown schematically in Fig. \ref{sfig:Methodtheoryfig}. 
First, quasiparticles can tunnel out of the island as either electrons or holes with a rate proportional to the $N_\text{QP}$ \cite{saira2012vanishing} (yellow arrows). As the quasiparticles are generated in pairs, no branch imbalance is created. Starting from $N=0$, tunneling can happen into both of the two states $N=\pm 1$ which leads to the factors $\sigma$, $\bar{\sigma}$ in Eq. (1) in Methods. 
Second, quasiparticles may in principle also tunnel back into the superconductor from the normal leads. We neglect the process $N=0 \rightarrow \pm 1$, $N_\text{QP} \rightarrow N_\text{QP} +1$ (backtunneling starting from $N=0$) due to its high energy cost (see discussion below), but include backtunneling starting from states with $N$ odd (red arrows). Third, we account for quasiparticle recombination with a rate scaling as $N_\text{QP}^2$ for $N_\text{QP} \geq 2$ (dotted green arrows), but we neglect the probability of another Cooper pair breaking event happening within a burst due to the observed orders of magnitude separation between the tunneling and Cooper-pair-breaking timescales. Finally, we also include two-electron Andreev tunneling between $N=\pm 1$ with a $N_\text{QP}$-independent rate. The prefactors of each of these rates are parameters of the model. 
We point out that the model is charge symmetric, that is, the rates  are unaffected by changing $N=\pm 1$ to $\mp 1$. While the model could readily be extended to charge asymmetric rates, this is not needed for the purposes in the present paper, as only the Andreev tunneling rates, which do not contribute to the quasiparticle dynamics, are expexted to be asymmetric in the experiment.  

To obtain the time evolution of $P(N_{\text{QP}},N,t)$ after a pair breaking event occuring at time $t_\text{PB}$, we take as initial condition the distribution of number of broken pairs $N_{\text{pair}}=N_{\text{QP}}/2$ determined experimentally (see Fig. 2 in the main text). This gives
\begin{equation}
P(N_{\text{QP}},N,t_\text{PB})=e^{-\lambda_\text{QP} N_{\text{QP}}}\left(e^{2\lambda_\text{QP}}-1\right)\delta_{N,0}, 
\label{eq:breakprob}
\end{equation}
for $N_{\text{QP}}=2,4,6..$ and zero otherwise, with $\lambda_\text{QP}=\lambda/2=0.45$. However, since only the electron tunneling events following the pair breaking event itself can be detected in our setup, in a number of cases we consider the distribution $P(N_{\text{QP}},N,t)$ after the first tunneling event occuring at $t=0$ (not introducing a separate notation, the distribution considered is clear from the context). The initial condition for this conditional probability distribution is given by
\begin{equation}
P(N_{\text{QP}},N,0)=e^{-\lambda_\text{QP} (N_{\text{QP}}+1)}\left(e^{2\lambda_\text{QP}}-1\right)\frac{\delta_{N,1}+\delta_{N,-1}}{2} 
\label{eq:tunnelprob}
\end{equation}
for $N_{\text{QP}}=1,3,5..$ and zero otherwise. This distribution accounts for the time evolution from the pair breaking to the first tunnel event. We point out that in obtaining Eq. (\ref{eq:tunnelprob}) from (\ref{eq:breakprob}), quasiparticle recombination is neglected. This is motivated by the observation that $\Gamma_{\text{QP}} > \Gamma_{\text{R}}N_{\text{QP}} $
for all relevant $N_{\text{QP}}$, as discussed below. 

From the probability distribution $P(N_{\text{QP}},N,t)$ after the first tunneling event occuring at $t=0$, the different quantities shown can be obtained. The formally infinite set of equations in (\ref{eq:rateeq}) is solved by truncating the quasiparticle number at some maximum $N_{\text{QP}}=N_{\text{QP}}^{\text{max}}$, such that the result is independent of $N_{\text{QP}}^{\text{max}}$. In the main text (including Methods) the basic model, considering only quasiparticle out-tunneling, is discussed. For the quantities analyzed in the Supplementary information, the theoretical burst length distribution, shown in Fig. \ref{sfig:burstlength}, is obtained by a Monte Carlo simulation of Eq. (\ref{eq:rateeq}). Moreover, the effective time-dependent single-electron rates $\Gamma_{\pm 1 \rightarrow 0}(t)$ and $\Gamma_{0 \rightarrow \pm 1}(t)$ are given by 
\begin{equation}
\Gamma_{N \rightarrow N \pm 1}(t)=\frac{\sum_{N_{\text{QP}}} P(N_{\text{QP}},N,t)\Gamma_{N_{\text{QP}} \rightarrow N_{\text{QP}}-1, N \rightarrow N \pm 1}}{\sum_{N_{\text{QP}}} P(N_{\text{QP}},N,t)}
 \end{equation}
where the sum runs over $N_\text{QP}=2,4,6,...$ for $N=0$ and $N_\text{QP}=1,3,5,...$ for $N=\pm 1$. The probability for the empty state $(0,0)$ is thus not included in the definition of the effective rates, in agreement with how the rates are obtained from the experimental data. Hence, the rates describe only the tunneling within the bursts. 

From $P(N_{\text{QP}},N,t)$ we can also obtain the time dependent probabilities for having a charge $N=0,\pm 1$ on the island. We focus on the total probabilities for having an odd ($N=\pm 1$) or an even ($N=0$) charge, defined as 
\begin{equation}
P(N=\pm1,t)=\sum_{N_{\text{QP}}=1,3,5..}\left[P(N_{\text{QP}},N=1,t)+P(N_{\text{QP}},N=-1,t)\right]    
\end{equation}
and
\begin{equation}
P(N=0,t)=\sum_{N_{\text{QP}}=2,4,..}P(N_{\text{QP}},N=0,t)    
\end{equation}
respectively. As for the effective rates, the probabilty for $N_{\text{QP}}=0$ is not included, in order to only describe the probabilities within the bursts.

\clearpage
\section{Supplementary Note 6: System characterization}

In this section, we discuss estimates for the various parameters contributing to the rates in our model considered in Supplementary Note 5 above, and show that in our experiment the dominant process after a Cooper-pair breaking event is the tunneling out of quasiparticles from the island, leading to the simple model used in the main text. 

\subsection{Charging energy, superconducting gap and electron temperature}
\label{sec:charging-energy}
\begin{figure}
    \includegraphics{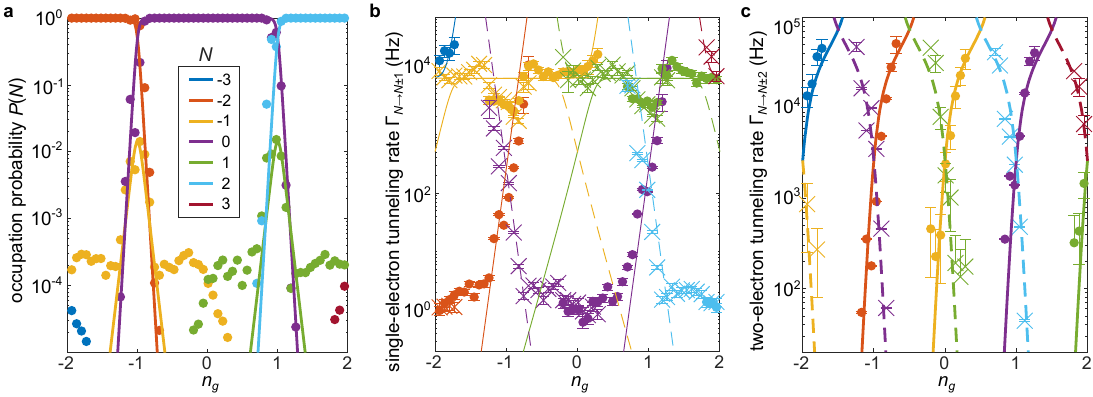}
    \caption{Measured \textbf{a} occupation probabilities $P(N)$, \textbf{b} single-electron tunneling rates $\Gamma_{N \rightarrow N\pm 1}$ and \textbf{c} two-electron tunneling rates $\Gamma_{N \rightarrow N\pm 2}$ as a function of the gate offset $n_g$. Symbols are experimental data, while lines are simulated results (see text). In panels \textbf{b} and \textbf{c}, dashed (solid) lines and crosses (circles) indicate transitions where $N$ decreases (increases). Error bars in \textbf{b} and \textbf{c} represent statistical uncertainty from the number of tunneling events measured per point. 
    }
    \label{sfig:gate-dependence}
\end{figure}

We determine the charging energy $E_C \approx 90$~$\mu$eV and superconducting gap $\Delta \approx 220~\mu$eV of the device by fitting the measured occupation probabilities of the charge states $N$ at elevated bath temperatures $T_{bath}$ between 120~mK and 215~mK, where we expect the device properties to be set by thermal excitations. The obtained values are consistent with the device geometry and typical gap values for thin aluminum films. 
Due to the charging energy $E_C \approx 0.4\Delta$, at our operating point where $n_g \approx 0$ the charge states with $|N|\geq 1$ carry a high energy cost $> \Delta$. In contrast, the charging energy cost for occupying the charge states $N=\pm1$ is less than $\Delta$ and they may be occupied for observable times if quasiparticles are present on the island. 
Hence the observed charge states relevant for the quasiparticle dynamics are $N=0,\pm 1$.

In Fig. \ref{sfig:gate-dependence}, we show the tunneling rates and occupation probabilities versus $n_g$, measured at the refrigerator base temperature $T=20$~mK in similar conditions as the data shown in Figs. 1-3 of the main text. Each experimental point here consists of averages over 30 s of data, with the tunneling rates calculated as the number of events divided by the time spent in the starting charge state.
The solid lines are simulations similar to those  in Refs.  \cite{mannila2019detecting, maisi2013excitation}. They incorporate the rates shown in Fig. \ref{sfig:Methodtheoryfig}, as well as thermally activated backtunneling of quasiparticles, and Andreev tunneling events  with rates given in Ref. \cite{averin2008nonadiabatic}. 
As Andreev rates are insensitive to the presence of quasiparticles in the superconductor in the few-quasiparticle regime considered here, the $n_g$ dependence of the Andreev rates allows fitting the electron temperature at the base temperature $T_N = 100$~mK. This temperature also reproduces the occupation probability of the state $N=1$ at $n_g=1$. Here, we do not include a Cooper pair breaking rate in the simulations, and thus the occupation probabilities and tunneling rates resulting from these nonequilibrium events around $n_g=0$ are not reproduced by our model. The observation of gate-independent single-electron tunneling rates over a large range of $n_g$, like in Ref. \cite{mannila2019detecting}, is a strong indication that the observed tunneling events are due to nonequilibrium Cooper pair breaking on the superconducting island.  For the data shown in Figs. 1-3 of the main text, $n_g \approx 0.05$. 
In the data shown in Fig. 4 of the main text, the burst rate is obtained from data where the gate offset is between $n_g = -0.2$ and $n_g = 0.2$ (see also Fig. \ref{sfig:burst-rate-tbath-ng} below).

\subsection{Single-quasiparticle tunneling rates}

\subsubsection{Quasiparticle tunneling out}
A theoretical value for the single-quasiparticle tunneling rate for removing a quasiparticle from the island can be estimated as $\Gamma_\text{QP, theory} = (2e^2 R_T V D(E_F))^{-1}$ \cite{saira2012vanishing}. 
The tunnel resistance of the two tunnel junctions of the superconducting island in series was measured at room temperature to be 520~k$\Omega$, and we estimate the low-temperature resistance of the two junctions in parallel, relevant in the experiment where no bias is applied, to be $R_T = 150$~k$\Omega$. This incorporates a typical 15\% increase in the resistance when cooling down. Given the literature value for the density of states at the Fermi level in the normal state for both spins $D(E_\text{F}) = 2.15 \times 10^{47}$ J$^{-1}$ m$^{-3}$~\cite{lerch2005quantum}, and the known island volume $V=$ 35 nm $\times$ 0.5 $\mu$m $\times$ 2 $\mu$m, we obtain $\Gamma_\text{QP, theory} \approx 17$~kHz. As literature values for the density of states differ by up to a factor of 1.5 \cite{ashcroftmermin,lerch2005quantum} and there is some uncertainty in the junction resistances as well, we find the estimated value in fair agreement with the measured $\Gamma_{\text{QP}}=8.0$~kHz. 

\subsubsection{Quasiparticle backtunneling}

The data shown in the main text was obtained at $n_g \approx 0.05$. At this operation point, the expected rate for the thermally activated transition $N=0 \rightarrow \pm 1$, $N_{\text{QP}} = 0 \rightarrow 1$ adding a quasiparticle into the island is negligible (below $10^{-6}$~Hz), and in any case, the rate of such events is bounded by the measured burst rate. 

The thermally activated rate of the transition 
$N=\pm1 \rightarrow 0$, $N_{\text{QP}} = 1 \rightarrow 2$ for a quasiparticle tunneling back from the leads from the energetically unfavorable charge states $N=\pm 1$ is on the order of 1 kHz or below. However, this rate is difficult to estimate precisely due to its exponential dependence on $E_C$, $\Delta$ and $n_g$, although the independence of measured quantities on $n_g$ over a relatively large range, shown below in Figs. \ref{sfig:burst-rate-tbath-ng}, \ref{sfig:num1e}, and \ref{sfig:burstlength}, also supports the conclusion that backtunneling from the leads is negligible. We have performed also simulations with a finite rate $\Gamma_\text{B}$. This rate can be expected to be independent of the $N_\text{QP}$, the number of quasiparticles already present, as we always have only very few excitations present. 

\begin{figure}
    \centering
    \includegraphics{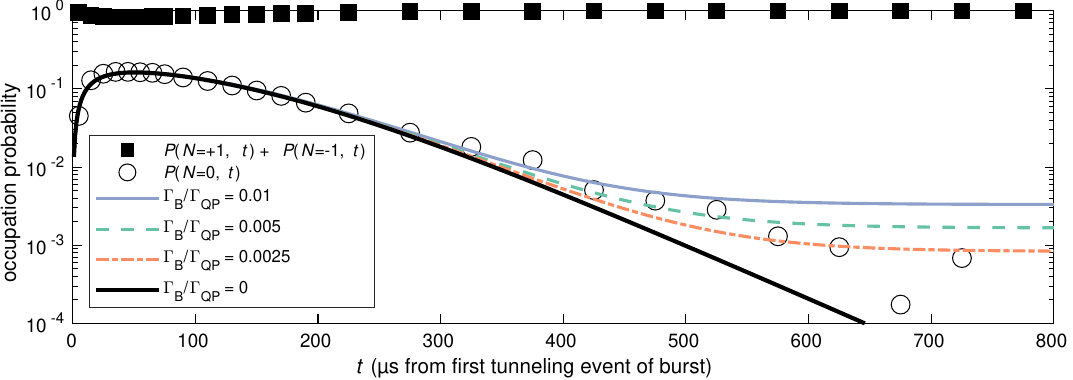}
    \caption{Occupation probabilities of the charge states $N$ as a function of time $t$ within the bursts of tunneling. Filled squares indicate the probability $P(N=+1,t)+P(N=-1,t)$, while open circles show the probability of $P(N=0,t)$. Lines are simulated probabilities $P(N=0,t)$ with varying backtunneling rates $\Gamma_{\text{B}}$. }
    \label{sfig:backtunneling}
\end{figure}

A sensitive probe of this backtunneling process are the time-dependent occupation probabilities $P(N,t)$ of the charge states within bursts of tunneling (in contrast with the probabilities $P(N_\text{QP},t)$ discussed in the main text). This can be intuitively be understood as follows: If there is no backtunneling, at long times from the initial pair-breaking event, a burst must either end or be in the charge state $\pm 1$. If there is a nonzero backtunneling rate, there would be a non-zero occupation probability for the state $N=0$ even at long times. Calculations show that the long-time limiting value is independent of the initial conditions (number of Cooper pairs broken) and given by $P(N=0) = (-3-q+\sqrt{(1+q)(9+q)})/2$, where $q = \Gamma_\text{B}/\Gamma_\text{QP}$ is the backtunneling rate normalized by the single-quasiparticle tunneling rate. 

In Fig. \ref{sfig:backtunneling}, we show the distribution $P(N=0,t)$ and $P(N=-1,t)+P(N=+1,t)$ obtained from the same dataset as used in Figs. 1-3 of the main text. Here, we show the total probability of the odd states $P(N=-1)+P(N=+1)$, as unequal Andreev tunneling rates in the experiment change the relative probabilities of the states $N=+1$ and $N=-1$, although they do not influence the quasiparticle dynamics. From the measured upper bound, we calculate $\Gamma_\text{B} < 0.01 \Gamma_\text{QP}$ and are thus justified in neglecting backtunneling. Hence, each observed single-electron tunneling event must correspond to an event removing a quasiparticle from the island, which enables reconstructing the instantaneous quasiparticle number. 

\subsection{Recombination rate}

We estimate the recombination rates based on the expression $\Gamma_\text{rec}(N_\text{QP}) = \Sigma \Delta^2 N_{\text{QP}}^2/ [12 \zeta(5) D(E_\text{F})^2 k_B^5 V] \equiv \Gamma_\text{R} N_{\text{QP}}^2$ given in Ref. \cite{maisi2013excitation}. Here, $\Sigma$ is the electron-phonon coupling constant in the normal state, $\zeta$ is the Riemann zeta function, and $k_B$ the Boltzmann constant. We use the measured values of the superconducting gap $\Delta \approx 220~\mu$eV and volume $V=2 \times 0.5 \times 0.035~\mu$m$^3$. With $\Sigma = 2 \times 10^8$~W~K$^{-5}$m$^{-3}$ consistent with literature values \cite{kautz1993selfheating,kauppinen1996electronphonon,meschke2004electron} and our recent measurements on similar Al films \cite{phonon-paper}, we obtain $\Gamma_\text{R}=25$~Hz and $N_\text{QP} \Gamma_\text{R} < \Gamma_\text{QP}$ up to $N_\text{QP} = 320$. Alternatively, the measured recombination constant $r=1/80$~ns in Ref. \cite{wang2014measurement}, connected to $\Gamma_\text{R} = r/(D(E_\text{F}) \Delta V)$, we obtain $\Gamma_\text{R} = 50$~Hz, although the effect of phonon trapping \cite{rothwarf1967measurement} is likely somewhat weaker in our experiment due to the thinner film and different substrate used. Hence we are justified in ignoring recombination in our basic model, shown in the main text.

Given these rates, we estimate the probability of a Cooper-pair-breaking event being undetected because the quasiparticles recombined before a tunneling event took place. An upper limit can be obtained by considering an event where only a single Cooper pair was broken, which leads to the shortest bursts. The relevant tunneling rate to compare with is the total tunneling rate out of the state $N=0$, $N_{\text{QP}}=2$, which is 
$4 \Gamma_\text{QP} = 32$~kHz. With $\Gamma_\text{R} = 25$~Hz,  in only 0.3\% of the cases the quasiparticles would recombine before tunneling out.
Even with the order of magnitude higher value of $\Sigma = 1.8 \times 10^9$~W~K$^{-5}$m$^{-3}$ in Ref. \cite{maisi2013excitation}, only 3\% of Cooper-pair breaking events would be undetected due to recombination. % 0.9/(32+0.9) = 2.7%. 0.1/(32+0.1) = 0.3% 
A non-zero recombination rate also reduces the count of single-electron tunneling events in a detected burst if some quasiparticles decay before tunneling out, meaning that in the presence of non-negligible recombination the true proportion of events which break more than one Cooper pair would be greater than 40\%.  

\subsection{Probability of undetected quasiparticles within the quiet periods}

In the quiet periods between the bursts of tunneling, quasiparticles may be present without leading to detected tunneling events, if the tunneling events were missed by our detector due to the finite bandwidth. We estimated above that the 1.4\% of bursts of tunneling that are shorter than 3~$\mu$s will be missed, and the probability that at any given moment quasiparticles would be present due to such an event is $\Gamma_{\text{burst}} \times 0.014 \times 3~\mu$s $= 6 \times 10^{-8}$.  Another possibility is that the quasiparticles recombined before tunneling out. Estimating cautiously that this happens after 1\% of the Cooper pair breaking events, the probability that quasiparticles were present due to such an event would be $\Gamma_{\text{burst}} \times 0.01 \times 30~\mu$s $= 5 \times 10^{-7}$. 

\clearpage
\section{Supplementary Note 7: Quasiparticle number distribution and single-quasiparticle rate versus time within cooldown, gate offset and bath temperature}

\begin{figure}
    \centering
    \includegraphics[width=\textwidth]{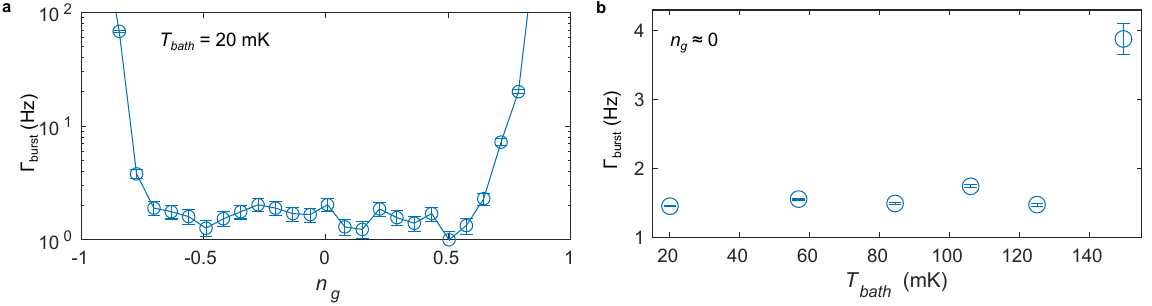}
    \caption{Burst rate shown as a function of \textbf{a} gate offset $n_g$, measured at a bath temperature of 20 mK, and \textbf{b} $T_{bath}$, measured at $n_g \approx 0$. All data shown here was measured between 125 and 140 days from the start of the first cooldown. The error bars in both panels reflect statistical uncertainty, but there may be systematical errors in the burst rate measured at higher temperatures due to changes in the charge detector performance. In panel \textbf{a}, a new burst is defined to start after $t_{sep}=250~\mu$s with no tunneling events in any charge state, whereas elsewhere we require $250~\mu$s in state $N=0$. Error bars represent statistical uncertainty.}
    \label{sfig:burst-rate-tbath-ng}
\end{figure}

In this section, we study the origin of the quasiparticles in further detail. In Fig. \ref{sfig:burst-rate-tbath-ng}, we show that the burst rate was independent of $n_g$ over a wide range, which is expected as the bursts are due to Cooper pair breaking events occurring independent of the state of the superconducting island. Near $n_g=1$, the burst rate increases due to thermal activation of single-electron and Andreev tunneling. The burst rate at $n_g \approx 0$ was also roughly independent of the refrigerator mixing chamber temperature up to 125 mK, which is expected as the Cooper pairs are broken by a non-thermal process. At 150 mK the measured burst rate again increases due to thermal activation of single-electron tunneling events.

\begin{figure}
    \centering
    \includegraphics[width=\textwidth]{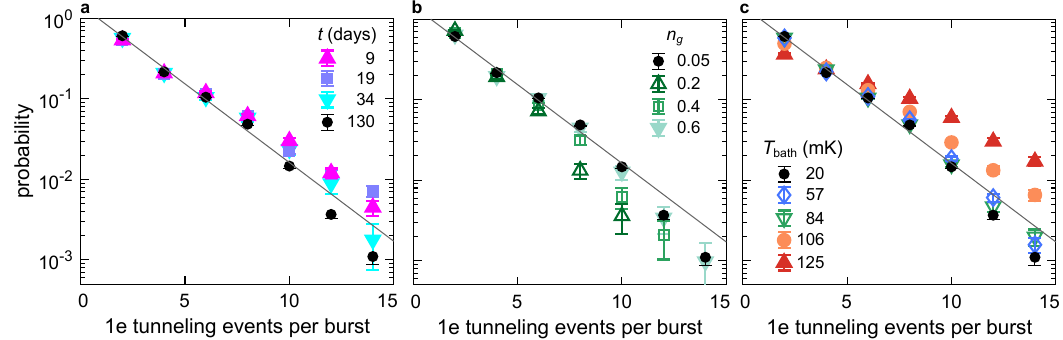}
    \caption{Distribution of single-electron tunneling events versus \textbf{a} time from start of the cooldown,  \textbf{b} gate offset, or \textbf{c} bath temperature,. The data in panels \textbf{b} and \textbf{c} was measured 125 to 140 days after the start of the first cooldown. In all panels, the black circles correspond to the dataset shown in the main text, while the solid line indicates exponential fits. Error bars represent statistical uncertainty.
    \label{sfig:num1e}
    }
\end{figure}

The distribution of single-electron tunneling events per burst was also unchanged as a function of time within cooldown, gate offset and bath temperature up to $T_{bath} = 85$~mK (Fig. \ref{sfig:num1e}). The independence on time within the cooldowns supports the conclusion that a single time-dependent quasiparticle source was dominant over the whole experimental period. 
At temperatures above 100 mK, the mean number of single-electron events per burst increases somewhat, which could be due to increased rates for thermally activated backtunneling (see above). They could also be due to noise interpreted as spurious tunneling events, since the charge detector performance deteoriates with increasing temperature. 

\begin{figure}
    \centering
    \includegraphics[width=\textwidth]{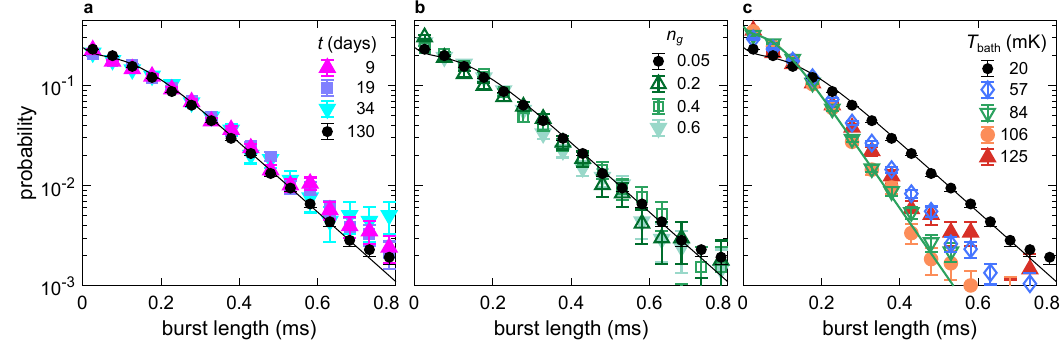}
    \caption{Burst length distributions versus \textbf{a} time from beginning of the second cooldown,  \textbf{b} gate offset, or \textbf{c} bath temperature,. The data in panels \textbf{b} and \textbf{c} was measured 124 to 140 days after the start of the first cooldown. Black circles in all panels correspond to the dataset shown in Figs. 1-3 in the main text, and the black solid line is the theoretical burst length distribution predicted by the basic model. In \textbf{c}, the green solid line is the burst length distribution calculated with a scaled single-quasiparticle tunneling rate 13 kHz, compared to 8 kHz used for the base-temperature data. Error bars represent statistical uncertainty.
    \label{sfig:burstlength}
    }
\end{figure}

To study the dependence of the single-quasiparticle tunneling rate $\Gamma_\text{QP}$ on these control parameters, as a complement to the quantities discussed in the main text we show in Fig. \ref{sfig:burstlength} the measured distribution of the burst lengths. This quantity requires less statistics to extract reliably than the effective rates shown in the main text, and our simulations show that after the initial 200 $\mu$s the distribution decays exponentially with the slope set by the single-quasiparticle rate. The initial deviation at $t<200~\mu$s from the exponential behaviour is due to the finite probability of more than one Cooper pair breaking and allows for an independent determination of the parameter $\lambda$ used as the initial condition for the simulations. We find that the single-quasiparticle rate is unchanged by the time during cooldown or $n_g$, as expected. At an increased $T_{bath}$, the single-quasiparticle rate increases by roughly a factor of 50\%, but the predictions of our simple model fit the measured burst length distribution exceedingly well when only this parameter is changed, as shown in Fig. \ref{sfig:burstlength}\textbf{c}. 

\clearpage
\section{Supplementary Note 8: Parity switching, quasiparticle tunneling and single-electron trapping}

\begin{figure}
    \centering
    \includegraphics{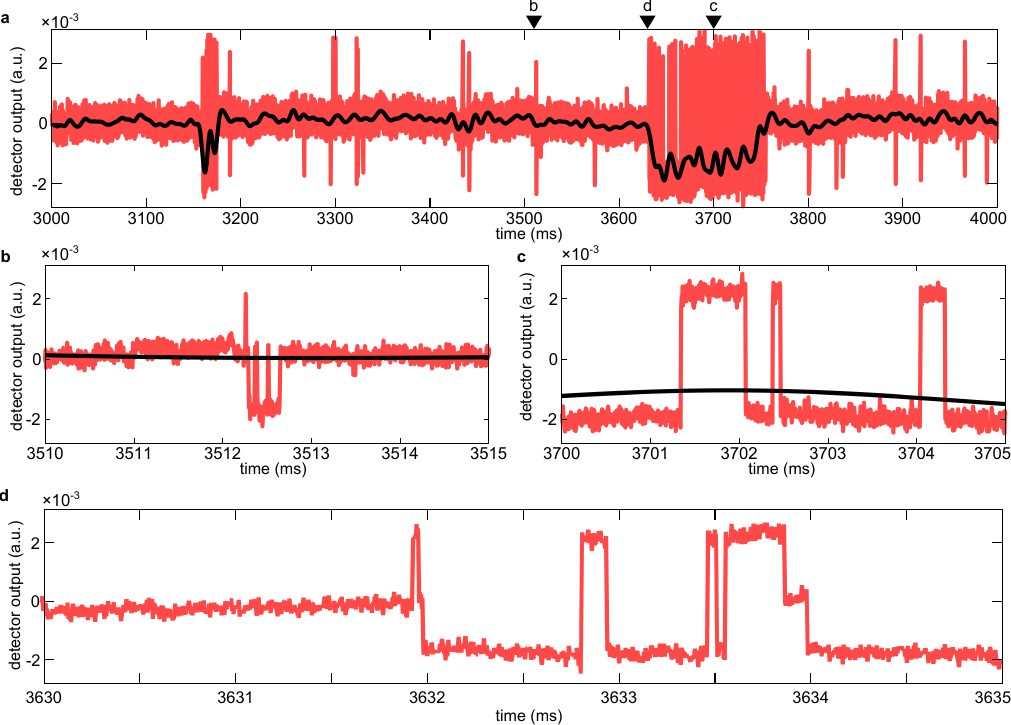}
    \caption{Portions of time trace from second cooldown, filtered with 150 kHz cutoff (red) and 100 Hz cutoff (black). Panels \textbf{b} and \textbf{c} show 5-ms portions of the trace in regimes (1) and (2) (see text), while panel \textbf{d} shows that in this case the transition between the two regimes happens via three single-electron tunneling events in rapid succession, visible at $t=3632~$ms. }
    \label{sfig:fluctuator}
\end{figure}

In the main text, we concentrate on the detection of quasiparticle-free periods and Cooper pair breaking events, which are followed by quasiparticles relaxing by tunneling from the superconducting island to the normal metal leads. In this section, we present in more detail the trapping effects mentioned in the second-to-last paragraph of the main text, and elaborate on the various timescales associated with parity switching on the superconducting island. We define "parity switch" as any event which changes the parity of $N$, the number of electrons on the superconducting island. The typical timescale between such events could be called a "parity lifetime". However, as our results show, there are multiple timescales and processes associated with parity switches.  

As presented in Fig. 1 of the main text, the typical dynamics measured at $n_g = 0$ is such that the system spends most of the time in the charge state $N=0$ and no quasiparticles are present. After a Cooper pair breaking event, the parity of $N$ changes several times during a short period of time as the quasiparticles tunnel out, but returns to its original value within a millisecond. The timescale of Cooper pair breaking events is the burst rate $\Gamma_{\text{burst}}$ (a few Hz), while the subsequent parity switches occur on a timescale related to the single-quasiparticle tunneling rate $\Gamma_{\text{QP}}$ (on the order of 10 kHz).

In the second cooldown with the same device we also detected events which change the tunneling dynamics at a rate on the order of 1 Hz.
Figure \ref{sfig:fluctuator} shows a typical example of these second type of events. Before $t$ = 3630  ms, the system shows the above discussed Cooper pair breaking dynamics, see panel \textbf{a} and the zoom-in in panel \textbf{b}. We denote this as regime (1). Then the operation switches at $t = 3630$~ms to regime (2) and subsequently switches back  at $t = 3750$~ms. As seen in the zoom-in, panel \textbf{c}, in this second regime, the system undergoes Andreev events between $N = \pm1$ and the state $N = 0$ is avoided. This observation demonstrates a second type of parity switching where the most probable charge state changes from even to odd or vice versa for a longer time than in th bursts of regime (1). 
At the beginnings and ends of these changes, in roughly half of the cases we observe several single-electron tunneling events within 1 ms, as visible in panel \textbf{d}. The switching rate between the two regimes, with dominating charge states $N=0$ and $N=\pm 1$ respectively, also decreased during the second cooldown by a similar factor as the burst rate shown in Fig. 4 over the same time period. 

We interpret these longer time scale parity switching events to arise from trapping and escaping of an electron in localized state in close proximity of the superconducting island. 
Such trapped states may be either subgap states in the superconductor or electronic states in impurities in close proximity or inside the superconductor. 

Interestingly, we observe that the switching to the second regime is triggered for roughly half of the cases by several single-electron events as shown in \ref{sfig:fluctuator}\textbf{d}. Our interpretation here is that initially, a Cooper pair breaking event (with the rate $\Gamma_{\text{burst}}$) triggers the burst shown in Fig. 1 of the main article. During the subsequent tunneling-out of the quasiparticles from the superconductor, one particle is trapped in the localized state, causing the switch to the second regime.   e. This interpretation is also supported by our observation that during the second cooldown, the switching rate to the second regime decreases over time from the cooldown start, with a similar factor as the burst rate. The switching rate also varied from tens of minutes or hours in the first cooldown to roughly 1 s in the second cooldown while the burst dynamics remained the same. This suggest that the trapped states form somewhat randomly during the cooldown while the burst dynamics are a robust feature of the quasiparticle excitation dynamics.

For acquiring the 4.6 h of data comprising the dataset used in Figs. 2 and 3 of the main text, we made sure that during the measurements no such switching to the regime (2) nor discrete jumps of $n_g$ took place. Even in this dataset, roughly 0.3\% of the measured bursts of tunneling were  longer than 1.5 ms, which should be exceedingly unlikely according to our model. Such long bursts might be also explained by quasiparticles becoming trapped in a localized state, from which they escape with a slower rate, which would also explain the slight decrease in the rate $\Gamma_{\pm 1 \rightarrow 0}$ after $t > 500~\mu$s. We emphasize that at $t=500~\mu$s, over 90\% of bursts have already ended so this effect is only a minor correction to the dynamics.

For the data shown in Fig. 4 of the main text and the data points corresponding to $t=9,19,$ or $34$ days in Figs. \ref{sfig:num1e} and \ref{sfig:burstlength}  measured in the second cooldown, we have only used the portions of the data where the dominant charge state is $N=0$. We have inferred this from the measured traces by low-pass filtering the data with a cutoff set to 100 Hz, so that the fast bursts of tunneling are filtered out. Such a filtered trace is shown also in Fig. \ref{sfig:fluctuator}(a,b,c). We then identify the dominant charge state by applying similar thresholds as in determining the charge states, described previously. 

%\bibliography{phonons}
%\bibliography{../../phonons/phonons}
%\input{qp_supplement.bbl}